\documentclass[12pt,a4paper]{article}
\usepackage[textheight=23cm,textwidth=15.5cm]{geometry}

\usepackage[utf8x]{inputenc}
\usepackage{graphicx}
\usepackage{cite}
\usepackage{here}
\usepackage{amsmath}
\usepackage{amssymb}

\usepackage{bm}

\title{The Polarizable Embedding Density Matrix Renormalization Group Method}
\author{Erik D. Hedeg{\aa}rd\thanks{erik.hedegard@phys.chem.ethz.ch} and Markus Reiher\thanks{markus.reiher@phys.chem.ethz.ch}}

\begin{document}

\maketitle

\hspace*{-0.9cm}

\begin{center}
	ETH Z\"urich, Laboratorium f{\"u}r Physikalische Chemie, \\ Vladimir-Prelog-Weg 2, 8093 Z\"urich, Switzerland. \\
\end{center}

\hspace*{0.5cm}

\begin{abstract}
The polarizable embedding (PE) approach is a flexible 
embedding model where a pre-selected region out of a larger system is described quantum mechanically while the interaction with the surrounding environment is modeled through an 
effective operator. 
This effective operator represents the environment by 
atom-centered multipoles and polarizabilities derived from quantum mechanical calculations on (fragments of) the environment. 
Thereby, the polarization of the environment is explicitly accounted for. 
Here, we present the coupling of the PE approach with the density matrix renormalization 
group (DMRG). This PE-DMRG method is particularly 
suitable for embedded subsystems that feature a dense manifold of frontier orbitals which requires large active spaces. Recovering such static electron-correlation effects 
in multiconfigurational electronic structure problems, while accounting for both electrostatics and polarization of a surrounding environment, allows us to 
describe strongly correlated electronic structures in complex molecular environments.  
We investigate various embedding potentials for the well-studied first excited state of water with active spaces that correspond to 
a full configuration-interaction treatment.  Moreover, we study the environment effect on the first excited state of a 
retinylidene Schiff base within a channelrhodopsin protein. For this system, we also investigate the effect of dynamical correlation 
included through short-range density functional theory.
\end{abstract}


\section{Introduction}
Many chemical reactions occur in a medium, e.g., in a solvent.  
Explicitly taking this environment into account  in super-molecular calculations dramatically increases
the computational cost for most quantum mechanical methods. 
However,  as  most chemical properties are local a so-called \textit{focused} model\cite{born1920,kirkwood1934,onsager1936,kirkwood1938} can be applied. 
In such a model, a pre-defined region, denoted the quantum mechanical (QM) region, is described by 
an accurate electronic structure method, whereas interactions with the remaining system are described through an effective operator.
Focused models have 
the additional advantage that the interpretation of the results obtained is kept simple due to the explicit system--environment separation. 

In focused models, 
the most widespread approximation for the environment is a dielectric continuum, where the environment is characterized solely 
by a dielectric constant.\cite{tomasi2005} 
Accordingly, continuum models can capture bulk effects of a solvent, but are less meaningful for structured environments 
with specific interactions such as hydrogen bonding. 
 A more general approach is to consider a discrete environment where all molecules are modeled 
explicitly. 
One realization of explicit models relies on a classical description of the environment, 
representing it by point charges. The most widely applied models employ partial charges derived from a 
molecular force field, as done in the quantum-mechanics molecular mechanics (QM/MM) hybrid models\cite{warshel1976,senn2009}. 
These schemes usually 
allow only the electron density of the QM system to be polarized. 
Whereas the QM system will also polarize the environment, this is normally neglected in QM/MM calculations. 

Various methods that include the environment polarization have been proposed (see, for instance, Refs.\ \citenum{slip10,defu11}).
However, some of them turn out to be computationally more demanding compared to their purely electrostatic counterparts. 
This is somewhat surprising as the polarizable methods are in fact closer to the spirit of the first QM/MM scheme by Warshel and Levitt\cite{warshel1976}. 
One option, that has been explored within Density Functional Theory (DFT), is a full quantum mechanical description with frozen parts in the 
environment through an effective potential. This is known as Frozen Density Embedding 
(FDE)\cite{senatore1986,cortona1991,wesolowski1993}. 
In FDE, polarization can be included by an iterative procedure known as freeze-and-thaw cycles\cite{wesolowski1996} 
in which the role of system and environment is interchanged for the subsystems until convergence is reached. 
The FDE method is not restricted to DFT and 
the embedded subsystem can also be described by a wave function 
which is usually denoted WFT-in-DFT\cite{govind1998,govind1999,manby2012,hoefner2013,csaba2013,csaba2014,dresselhaus2015,hoefner2016}. 
WFT-in-WFT embedding schemes have also been investigated, e.g.,~by Chan and co-workers\cite{knizia2012,knizia2013,wout16},
by Scuseria and co-workers\cite{scus14,scus14b}, and by Fromager and co-workers\cite{fromager2015,from16}. 

\textit{Polarizable Embedding}\cite{soderhjelm2009,olsen2010,olsen2010b,sneskov2011,schwabe2011,list2013,hedegaard2013a,hedegaard2014,hedegaard2015a} (PE) 
relies on a different strategy, namely on atom-centered polarizabilities to represents the fragments in the environment.  
Contrary to other methods of this kind\cite{thompson1995,thompson1996,gao1997,jensen2003a,jensen2003b,yu2005}, 
PE employs both high-order multipoles and polarizabilities obtained from QM calculations, which makes the 
PE potential easily customized to a specific environment.

For locally excited states embedded in an environment, most studies apply some variant of time-dependent DFT which is based on response theory. 
Here, we instead aim at a state-specific \textit{ab initio} approach. The state-specific approach 
features aspects that are more appealing than methods based on response theory.
This was shown\cite{corni2005,cammi2005} by comparing the 
two approaches to the analytical solution for a four-state model (in a continuum solvation model). The two methods were shown 
to have different formal expressions for excitation energies under inclusion of a polarizable environment. In particular, the linear response 
approach fails to account for differences of the dipole moments in ground and excited states \cite{corni2005}. 
Yet, state-specific approaches have been dormant in this field for a long time. The interest was recently 
revived in studies on solvated systems with the FDE scheme coupled to a 
Quantum Monte Carlo wave function description for embedded system.\cite{csaba2013,csaba2014} 
In state-specific excited state optimizations, the treatment of static correlation effects is of high importance 
as excited states typically display large static correlation effects. 
One of the most capable methods to recover static correlation is the density matrix 
renormalization group (DMRG) algorithm\cite{white1992b,white1993,ors_springer,chan2008a,marti2010b,chan2011,marti2011,wouters2014rev,kurashige2014,szalay2015,yanai2015,garnet16}. 
Our group recently reported 
the implementation of an efficient second-generation DMRG program\cite{keller2014,dolfi2014,keller2015,knecht2016,keller2016} which 
relies entirely on matrix product operators. 
In this paper, we describe the coupling of this DMRG program with a PE scheme. 
This approach supplements our recently described coupling of DMRG
to the FDE scheme\cite{dresselhaus2015}. 

As an active-space method, DMRG relies on the selection of a proper orbital space from frontier molecular orbitals \cite{stein16a,stein16b}.
As a consequence, dynamical electron correlation must be considered either {\it a posteriori},
for instance by perturbation theory\cite{yana11,kurashige2014-cupt2,sharm14,soko16,guo16,knecht2016,roem16,wout16b} (diagonalize-and-then-perturb\cite{shavitt2002}), 
or {\it a priori} from the outset (in a perturb-and-then-diagonalize approach\cite{shavitt2002}). We recently 
investigated the latter option for DMRG\cite{hedegaard2015b} by employing a range-separated Hamiltonian\cite{savinbook} that recovers dynamical correlation 
through DFT by short-range (sr) density functionals (DMRG--srDFT)
in close analogy to the MCSCF--srDFT ansatz\cite{fromager2007,fromager2009,stoyanova2013}. 
The long-range part is then described by a DMRG wave function \textit{ansatz}. 
Herein, we also report the extension of our PE-DMRG approach to PE-DMRG--srDFT.   
 
This paper is organized as follows: In Section \ref{theory} we outline DMRG and PE theories as well as the 
required extensions for PE-DMRG to accommodate short-range functionals. 
Implementational aspects are described in Section \ref{implementation_details} and computational details in Section \ref{compmet}. 
We then proceed with applications of PE-DMRG and PE-DMRG-srDFT in Section \ref{results} that focus on photoexcitation of water
and the retinalidyne chromophore in the channelrhodopsin protein.  
Finally, conclusions and an outlook are given in Section \ref{conclusion}.

\section{Theory} \label{theory}

\subsection{The Complete-Active-Space Method} \label{CAS}

In this paper, we generally work in Hartree atomic units unless otherwise noted. 
We focus first on an isolated system in state $\alpha$ with an energy $E^{\rm iso}_{\alpha}$ defined as
\begin{equation}
 E^{\rm iso}_{\alpha} = \langle \Psi_{\alpha} \vert \hat{H}^{\rm iso}\vert \Psi_{\alpha} \rangle = \sum_{pq}h_{pq}D^{\alpha}_{pq} 
+ \frac{1}{2}\sum_{pqrs}g_{pqrs}P^{\alpha}_{pqrs} + {V}_{\text{nn}},
\label{E_iso}
\end{equation}
where $V_{\text{nn}}$ is the nuclear repulsion potential energy and $ h_{pq} $ and $g_{pqrs}$ are the usual one- and two-electron 
integrals over molecular orbitals $\phi_p (\bm{r})$, respectively. Orbital indices $p,q,r,s$ denote general 
spatial orbitals, $i,j,k,l$ inactive (doubly occupied) orbitals, and $u,v,x,y$ 
active (partially occupied) orbitals\cite{roos1980b,siegbahn1981}. 
 The system is described through an electronic wave function 
$\vert\Psi_{\alpha}\rangle$ for the $\alpha$-th (electronic) state; $\bm{D}^{\alpha} = \{ D^{\alpha}_{pq}\}$ and $\bm{P}^{\alpha}= \{ P^{\alpha}_{pqrs}\}$ are the corresponding 
one- and two-electron reduced density matrices (1-RDM and 2-RDM, respectively), 
\begin{align}
 D^{\alpha}_{pq}    & = \langle \Psi_{\alpha} \vert \hat{E}_{pq}\vert \Psi_{\alpha}  \rangle ,  \label{one-and-two-electron-dens-1} \\
  P^{\alpha}_{pqrs} & = \langle \Psi_{\alpha}  \vert \hat{e}_{pqrs}\vert \Psi_{\alpha}  \rangle . \label{one-and-two-electron-dens-2} 
\end{align}
The operator $\hat{E}_{pq}$ is defined as 
\begin{align}
 \hat{E}_{pq}   =  \hat{E}^{\uparrow}_{pq} +  \hat{E}^{\downarrow}_{pq} 
=  \hat{a}^{\dagger}_{p\uparrow}\hat{a}_{q\uparrow} +  \hat{a}^{\dagger}_{p\downarrow}\hat{a}_{q\downarrow}  , \label{E_pq}
\end{align}
where $\hat{a}^{\dagger}_{p}$ and $\hat{a}_{p}$ are creation and annihilation operators\cite{helgaker2004}, respectively, 
defined for orbital $\phi_p(\bm{r})$ with spin-up ($\uparrow$) and spin-down ($\downarrow$) quantum numbers.
The operator $\hat{e}_{pqrs}$ then reads
\begin{equation}
\hat{e}_{pqrs} = \hat{E}_{pq}\hat{E}_{rs} - \hat{E}_{ps}\delta_{qr} . \label{e_pqrs}
\end{equation}
All complete active space methods divide the Hamiltonian of the isolated system into active (``A'') and inactive (``I'') parts
\begin{equation}
 \hat{H}^{\rm iso} = \hat{H}^{\rm iso}_{I} + \hat{H}^{\rm iso}_{A}, 
\label{Hamilton_CAS_iso}
\end{equation}
with 
\begin{align}
 \hat{H}^{\rm iso}_{I} &= \frac{1}{2}\sum_{ij}\bigl(h_{ij} + f^{I}_{ij} \bigr)\hat{E}_{ij} + V_{\rm nn} \label{H_inactive} 
\end{align}
and 
\begin{align}
  \hat{H}^{\rm iso}_{A} &= \sum_{uv}f^{I}_{uv}\hat{E}_{uv} + \frac{1}{2}\sum_{uvxy}g_{uvxy}\hat{e}_{uvxy},   \label{H_active}
\end{align}
where $f^{I}_{pq}$ denotes an element of the inactive Fock matrix 
\begin{equation}
 f^{I}_{pq} = h_{pq} + \sum_{k}\bigl(2 g_{pqkk} - g_{pkqk} \bigr) \label{Fock_inact} . 
\end{equation}
Accordingly, the complete-active-space configuration-interaction (CAS-CI) energy is then given as a sum of an inactive energy, $E^{\rm iso}_{I}$, 
and an active energy, $E^{\rm iso}_{A}$: 
\begin{align}
E^{\text{iso}}_{\alpha} = 
\langle\Psi_{\alpha}\vert \hat{H}^{\rm iso}_{I}\vert \Psi_{\alpha}\rangle + \langle\Psi_{\alpha}\vert \hat{H}^{\rm iso}_{A}\vert \Psi_{\alpha}\rangle
 = E^{\rm iso}_{I,\alpha} + E^{\rm iso}_{A,\alpha} , \label{CAS-CI-energy} 
\end{align}
where
\begin{align}
E^{\rm iso}_{I,\alpha} & =  \frac{1}{2}\sum_{ij}\bigl(h_{ij} + f^{I}_{ij}\bigr) D^{I,\alpha}_{ij} + V_{\text{nn}} 
 = \sum_{i}\bigl(h_{ii} + f^{I}_{ii}\bigr) + V_{\text{nn}} ,  \label{inact-energy}
\end{align}
and 
\begin{align}
E^{\rm iso}_{A,\alpha}   & = \sum_{uv}f^{I}_{uv}D^{A,\alpha}_{uv} + \frac{1}{2}\sum_{uvxy}g_{uvxy}P^{A,\alpha}_{uvxy} \label{act-energy} . 
\end{align}
Note that we keep the $\alpha$ state index for the inactive part in order to emphasize its implicit dependence on the choice and type of inactive
orbitals.
In Eqs.~\eqref{inact-energy} and \eqref{act-energy} we have divided the 1-RDM into an inactive part 
$\bm{D}^{I,\alpha} = \{ D^{I,\alpha}_{ij}\} = \{ 2 \delta_{ij}\}$, 
and an active part, $\bm{D}^{A,\alpha} = \{ D^{A,\alpha}_{uv}\}$. In the following subsection, we discuss how the operators 
from Eqs.~\eqref{E_pq} and \eqref{e_pqrs} are constructed 
in a matrix-product-operator (MPO) based DMRG algorithm and how the (active) 1-RDM is obtained.  

\subsection{DMRG with Matrix Product Operators} \label{CAS_MPS_MPO_DMRG}

For the DMRG algorithm, the orbitals in the CAS are arranged on a linear \textit{lattice}, each orbital defining a \textit{site}. 
For a lattice of length $L$ (i.e., for a CAS with $L$ orbitals) the DMRG wave function can be written in the MPS formalism as\cite{rommer1997} 
%\begin{equation}
%\vert \Psi_{\alpha}\rangle = \sum_{\bm{\sigma}}c_{\bm{\sigma},\alpha} \vert \bm{\sigma}\rangle = 
% \sum_{\bm{\sigma}}M^{\sigma_1}_{\alpha} M^{\sigma_2}_{\alpha} \cdots M^{\sigma_L}_{\alpha}\vert \bm{\sigma}\rangle \label{MPS_state}.  
%\end{equation}
\begin{equation}
\vert \Psi_{\alpha}\rangle  = 
 \sum_{\bm{\sigma}}M^{\sigma_1}_{\alpha} M^{\sigma_2}_{\alpha} \cdots M^{\sigma_L}_{\alpha}\vert \bm{\sigma}\rangle \label{MPS_state}.  
\end{equation}
The (site-) matrices  $M^{\sigma_l}_{\alpha}$ in Eq.~\eqref{MPS_state}, defined for each site $l$, multiply to yield a CI coefficient 
($M^{\sigma_1}_{\alpha}$ and $M^{\sigma_L}_{\alpha}$ are vectors). $\vert\bm{\sigma}\rangle = \vert \sigma_1 , \sigma_2, \ldots,\sigma_L \rangle $ 
denotes an occupation number vector and 
on each site, four \textit{local} states ranging from doubly occupied to empty, 
$\vert\sigma_l \rangle = \bigl(\vert\hspace{-1.0mm}\uparrow\downarrow\rangle, \vert\hspace{-1.0mm}\uparrow\rangle, \vert\hspace{-1.0mm}\downarrow\rangle,\vert 0\rangle\bigr)$,  
are defined. The DMRG algorithm optimizes the site matrices iteratively. 

The MPS formalism can be transferred to operators yielding the {Matrix Product Operators} \cite{mcclloch2007,crosswhite2008}.
Our DMRG implementation employs an MPO form for all operators\cite{dolfi2014,keller2014,keller2015}. Accordingly, an operator $\mathcal{\hat{W}}$ will be 
 of the form  
\begin{equation}
\hat{\mathcal{W}} = \sum_{\bm{\sigma}\bm{\sigma}'}W_{\bm{\sigma}\bm{\sigma}'}\vert\bm{\sigma}\rangle\langle\bm{\sigma}'\vert ,  \label{mpo_1}
\end{equation}
where $W_{\bm{\sigma}\bm{\sigma}'}$ is given by
\begin{equation}
 W_{\bm{\sigma}\bm{\sigma}'} = \sum_{b_1,\cdots b_{L-1}}W^{\sigma_1 ,\sigma'_1 }_{1 b_1}\cdots W^{\sigma_{l} ,\sigma'_l }_{b_{l-1} b_{l}} \cdots 
W^{\sigma_{L} ,\sigma'_L }_{b_{L-1}1} ,  \label{mpo_2}
\end{equation}
so that an expectation value reads (see Refs.~\citenum{keller2015} and \citenum{schollwock2011} for details)
\begin{align}
  \langle\Psi_{\alpha}\vert \hat{\mathcal{W}}\vert \Psi_{\alpha}\rangle 
= & \sum_{\sigma_{L}\sigma'_L}\sum_{b_{L-1}}M^{\sigma_L \dagger }_{\alpha} W^{\sigma_{L}\sigma'_L}_{b_{L-1} 1 }\Bigl( 
\cdots \sum_{\sigma_2 \sigma'_2}\sum_{b_1}M^{\sigma_2 \dagger }_{\alpha} W^{\sigma_{2}\sigma'_2}_{b_{1} b_2 }  \notag \\
& \cdot \Bigl( \sum_{\sigma_1 \sigma'_1}M^{\sigma_1 \dagger }_{\alpha} W^{\sigma_{1}\sigma'_1}_{1 b_{1}  } M^{\sigma'_1 }_{\alpha}\Bigr) 
  M^{\sigma'_2 }_{\alpha} \cdots  \Bigr) M^{\sigma'_L }_{\alpha} \label{mpo_mps_2} . 
\end{align}
For the active part of the 1-RDM associated with $\hat{E}^{\uparrow}_{uv}$, we find 
\begin{align}
 D^{\uparrow\alpha}_{uv} = & \langle\Psi_{\alpha}\vert \hat{E}^{\uparrow}_{uv}\vert \Psi_{\alpha}\rangle \notag \\
= & \sum_{\sigma_{L}\sigma'_L}\sum_{b_{L-1}}M^{\sigma_L \dagger }_{\alpha} \mathbb{I}^{\sigma_{L}\sigma'_L}_{b_{L-1} 1 }\Bigl( 
\cdots \Bigl(\sum_{\sigma_v \sigma'_v}\sum_{b_{v-1}}M^{\sigma_v \dagger }_{\alpha} \mathbb{D}^{\sigma_{v}\sigma'_v  \uparrow}_{b_{v-1} b_v }  \cdots 
\Bigl(\sum_{\sigma_u \sigma'_u}\sum_{b_{u-1}}M^{\sigma_u \dagger }_{\alpha} \mathbb{D}^{\sigma_{u}\sigma'_u  \uparrow}_{b_{u-1} b_u } \notag \\
& \cdots 
 \sum_{\sigma_2 \sigma'_2}\sum_{b_1}M^{\sigma_2 \dagger }_{\alpha} \mathbb{I}^{\sigma_{2}\sigma'_2}_{b_{1} b_2 }  
 \cdot \Bigl( \sum_{\sigma_1 \sigma'_1}M^{\sigma_1 \dagger }_{\alpha} \mathbb{I}^{\sigma_{1}\sigma'_1}_{1 b_{1}  } M^{\sigma'_1 }_{\alpha}\Bigr) 
  M^{\sigma'_2 \dagger }_{\alpha} \cdots \Bigr)M^{\sigma'_u \dagger }_{\alpha} \cdots\Bigr)M^{\sigma'_v \dagger }_{\alpha}
 \cdots  \Bigr) M^{\sigma'_L \dagger }_{\alpha}  , \label{D_uv_MPO}
\end{align}
where $\mathbb{I}^{\sigma_{i}\sigma'_{i}}$  denotes unit matrices. 
Note that we have not explicitly accounted for the fermionic anti-commutation of the creation and annihilation operators in Eqs.~\eqref{D_uv_MPO}.  
As explained in Ref.~\citenum{keller2015}, this can be done by multiplying the matrix representations 
of the creation and annihilation operators with a 4$\times$4 \textit{fill matrix} related to a Jordan-Wigner transformation.  

\subsection{\label{theorya}Polarizable Embedding}\label{PE_scheme}

We  now consider the energy, $E^{\rm pe}$, associated with the interaction of environment and embedded system.  
 We will, in general, have 
 both electrostatic and polarization contributions, $E^{\rm es}_{\alpha}$ and $E^{\rm pol}_{\alpha}$, respectively,
\begin{equation}
 E^{\rm pe}_{\alpha} = E^{\rm es}_{\alpha} + E^{\rm pol}_{\alpha}  \label{PE_energy}. 
\end{equation} 
The PE scheme\cite{soderhjelm2009,olsen2010} divides the environment into (molecular) fragments. 
We associate to each of these fragments a set of electrostatic multipoles, usually
localized at the atomic centers\cite{gagliardi2004}. A set of 
charges, $q (\bm{r}_{s'})$, dipole moments $\mu_{\alpha}(\bm{r}_{s'})$, quadrupole moments $ Q_{\alpha\beta}(\bm{r}_{s'})$ etc., 
are in this way defined for each center, $s'$, in the environment (the prime distinguishes environment centers from orbital indices). 
In addition, the PE scheme allows for a polarization of the environment 
 by defining a set of localized polarizabilities $\{\bm{\alpha}_{s'} \}$ on each atomic center of the environment. 
We obtain both the localized multipoles and polarizabilities from a quantum chemical calculation (usually within DFT).
 $E^{\text{es}}_{\alpha}$ is therefore given by 
\begin{equation}
 E^{\text{es}}_{\alpha} = \langle\Psi_{\alpha}\vert \hat{V}^{\text{es}}\vert\Psi_{\alpha}\rangle + E^{\text{mul}},  \label{esenergy} 
\end{equation}
where $E^{\text{mul}}$ contains the 
%(constant) \erik{``Constant'' means that for a given environment structure/configuration, $ E^{\text{mul}}$ only depends on the environment atoms's coordinates and the pre-calculated multipole moments). However, the term with $ \langle\Psi_{\alpha}\vert \hat{V}^{\text{es}}\vert\Psi_{\alpha}\rangle$ depends on the wave function and therefore changes during an optimization (although the change is linear in the 1-RDM and therefore trivially calculated)} 
interaction of all multipoles in the environment. 
The operator $\hat{V}^{\text{es}}$ describes the interaction of all electrons in the QM region with the environment,   
\begin{equation}
 \hat{V}^{\rm es} = \sum_{s'}\left( E^{\text{es}}_{\text{n},s'}  +  \sum_{pq}V_{pq,s'}\hat{E}_{pq} \right) , \label{V_es_operator} 
\end{equation}
where the nuclear part, $E^{\text{es}}_{\text{n},s'}$ can simply be added to the electronic part. 
 $V_{pq,s'}$ are the electrostatic potential integrals for center $s'$ defined through a multipole expansion\cite{olsen2010,list2013,hedegaard2013a}
\begin{equation}
 V_{pq,s'} = \sum_{|k|=0}\frac{(-1)^{|k|}}{k!}t^{(k)}_{pq,s'}M^{(k)}_{s'} \label{multipole_exp} . 
\end{equation}
The multipole expansion in Eq.~\eqref{multipole_exp} is described through the composite index $k$ 
collecting $(k_x , k_y, k_z)$ in such a way that $k! = k_x ! k_y ! k_z !$ and $|k| = k_x + k_y + k_z$. In this notation, 
the $t^{(k)}_{pq,s'}$ in Eq.~\eqref{multipole_exp} are the integrals over the $T^{(k)}_{s'}$ 
interaction operators\cite{buckingham1967,stone2002} of order $k$
\begin{equation}
 t^{(k)}_{pq,s'} = -\langle \phi_p \vert T^{(k)}_{s'}\vert \phi_q \rangle ,  
\label{t_k}
\end{equation}
and the $M^{(k)}_{s'}$ are multipoles of order $k$ at center $s'$. 
The composite-index notation is detailed in Refs.~\citenum{olsen2010,olsen2011,hedegaard2013a} where also 
the explicit form of the $T^{(k)}_{s'}$ is given.

%\erik{Note on nomenclature: It is probably more correct to denote $\bm{F}$ as ``field strength'' (as introduced now). Field-strength is often denoted $\bm{E}$ in physics literature, but for some reason not in previous (polarizable) embedding papers. I have kept $\bm{F}$ to  match with previous PE papers}  
The  energy associated with polarization, $E^{\text{pol}}_{\alpha}$, is given by
\begin{equation}
 E^{\text{pol}}_{\alpha} = -\frac{1}{2}\bm{\mu}_{\text{ind}} \langle\Psi_{\alpha}\vert\hat{\bm{F}}\vert\Psi_{\alpha}\rangle ,
\end{equation}
where $\bm{\mu}_{\text{ind}}$ collects all dipole moments and $\hat{\bm{F}}$ all field-strength operators
induced locally at each center. 
For a system with $s'_{\text{tot}}$ \textit{polarizable} sites in the environment, we then have
\begin{align}
 \bm{\mu}_{\text{ind}} & = \bigl(\bm{\mu}^{\text{ind}}_1 , \bm{\mu}^{\text{ind}}_2, \ldots, \bm{\mu}^{\text{ind}}_{s'_{\text{tot}}} \bigr)^{T},\\
 \bm{\hat{F}} & = \bigl(\bm{\hat{F}}_1 , \bm{\hat{F}}_2, \ldots, \bm{\hat{F}}_{s'_{\text{tot}}} \bigr)^{T} . 
\end{align}
Note that   centers within the same fragment are not allowed to polarize 
each other and hence, the number of polarizable centers will be equal or less than the number of total centers in the environment.  
%The field-operator $\bm{\hat{F}} $ contains the field on each site in the environment. % As can be seen from Eq.~\eqref{inducedenergy} this means that
On the atomic center $s'$, $\bm{\hat{F}}_{s'}$ will have contributions from the multipoles ($\bm{F}^{\text{es}}_{s'}$) as well as 
from the nuclei ($\bm{F}^{\text{n}}_{s'}$) and electrons ($\bm{F}^{\text{e}}_{s'}$) in the QM region, 
\begin{equation}
 \bm{\hat{F}}_{s'} =  \bm{F}^{\text{es}}_{s'} + \bm{F}^{\text{n}}_{s'} + \hat{\bm{F}}^{\text{e}}_{s'}. \label{total_field}
\end{equation}
The two latter operators are defined as 
\begin{align}
 \bm{F}^{\text{n}}_{s'} & =  \sum_{M} Z_{M}\bm{T}^{(1)}_{M s'}  = \sum_{M}\frac{Z_{M}(\bm{R}_{M} - \bm{r}_{s'})}{|\bm{R}_{M} - \bm{r}_{s'}|^{3}} 
\end{align}
and
\begin{align}
 \bm{\hat{F}}^{\text{e}}_{s'} & = -\sum_{pq}\bm{t}^{(1)}_{pq,s'} \hat{E}_{pq} 
 = \sum_{pq} \left\langle\phi_p\left| \left(\frac{\bm{r}_{} - 
\bm{r}_{s'}}{|\bm{r}_{} - \bm{r}_{s'}|^{3}} \right)\right| \phi_q \right\rangle \hat{E}_{pq}  \notag 
  = - \sum_{pq}\bm{F}^{\rm e}_{pq,s'}\hat{E}_{pq}
\end{align}
where $Z_M$ and $\bm{R}_M$ are charge numbers and coordinates of the nuclei in the QM region, respectively. $\bm{r}$ is an electronic coordinate 
and $\bm{r}_{s'}$ the coordinates of the point-polarizability centers. 
%\erik{$\bm{F}^{\rm e}_{pq,s'}$ is not strictly necessary, but for aesthetic reasons i like these operators better for the definitions in Sec. 3. To make the connection between earlier PE papers (and also connect Eq.~\eqref{t_k} to the expression above) easy to see I have kept $\bm{t}^{(1)}$. } 

The dipole moment, $\bm{\mu}^{\text{ind}}_{s'}$, induced at atomic center $s'$ depends on the 
polarizability on that center, $\bm{\alpha}_{s'}$, and on the locally induced field strength, $\bm{F}_{s'}$, 
calculated as the observable over the operator given in Eq.~\eqref{total_field}.  However, it also depends on the induced dipole moments of the other sites, $\bm{\mu}_{\text{ind},s''}$, 
which can be accounted for by the dipole tensors, $\bm{T}^{(2)}_{s's''}$, given in Ref.\ \citenum{applequist1972},
\begin{align}
 \bm{\mu}_{\text{ind},s'} = \bm{\alpha}_{s'}\bm{F}_{s'} - \bm{\alpha}_{s'}\sum_{\substack{s'=1 \\ s''\neq s'}}\bm{T}^{(2)}_{s's''}\bm{\mu}^{\rm ind}_{s''} . 
\label{induced_dipoles}
\end{align}
Apart from being applied in QM/MM with polarizable force fields\cite{thompson1995,thompson1996,gao1997}, 
this expression can be found in the context of the classical description of molecular properties, 
e.g., in the work of Thole\cite{thole1981} and Applequist\cite{applequist1972,applequist1977} and 
earlier of Mortensen\cite{mortensen1968} and Silberstein\cite{silberstein1917a,silberstein1917b,silberstein1917c}. 
The induced dipole moments are usually obtained by re-writing Eq.~\eqref{induced_dipoles} into a matrix equation\cite{applequist1972}, 
\begin{equation}
\bm{\mu}_{\text{ind}} = \bm{R}\bm{F} , 
\end{equation}
where $\bm{R}$ is an inverted (super) matrix with $\bm{\alpha}^{-1}_{s'}$ tensors on the diagonal and (negative) dipole
tensors, $-\bm{T}^{(2)}_{s's''}$, as off-diagonal elements (see, e.g., Refs.~\citenum{jensen2003a,olsen2010,hedegaard2013a}).
The energy due to polarization in the environment can then be written as
\begin{align}
 E^{\text{pol}}_{\alpha} = -\frac{1}{2} \langle\Psi_{\alpha}\vert \bm{\hat{F}}\vert \Psi_{\alpha}\rangle^T \bm{R} \langle\Psi_{\alpha} \vert\bm{\hat{F}}\vert \Psi_{\alpha}\rangle  . \label{inducedenergy_2}
\end{align}
Following Ref.~\citenum{list2013}, Eqs.~\eqref{esenergy} and \eqref{inducedenergy_2} can be combined to yield the total energy as %the energy functional 
\begin{align}
 E_{\alpha}  = \langle\Psi_{\alpha}\vert \hat{H}^{\rm iso}\vert\Psi_{\alpha}\rangle + \langle\Psi_{\alpha}\vert \hat{V}^{\rm es}\vert \Psi_{\alpha}\rangle 
  -\frac{1}{2} \langle\Psi_{\alpha}\vert \bm{\hat{F}}\vert \Psi_{\alpha}\rangle^T \bm{R} \langle\Psi_{\alpha} \vert\bm{\hat{F}}\vert \Psi_{\alpha}\rangle + E^{\text{mul}} .     
\label{total_energy}
\end{align}
A solution for the active system can be obtained by a linear variation in the wave function parameters  
in Eq.~\eqref{total_energy}, corresponding to optimizing the pseudo-energy %(assuming $\langle\Psi |\delta\Psi\rangle = \langle\delta\Psi |\Psi\rangle =0$)
\begin{align}
 \mathcal{E}_{\alpha} = \langle\Psi_{\alpha}\vert \hat{H}^{\text{iso}} + \hat{V}^{\text{pe}} \vert\Psi_{\alpha}\rangle, 
\label{pseudo-energy}
\end{align}
where we defined the effective PE operator 
%\erik{this is how the expression of the PE operator is obtained - and why $E^{\text{mul}}$ that does not depend on wave function parameters disappear}
\begin{equation}
 \hat{V}^{\rm pe}   = \hat{V}^{\rm es} - \langle\Psi_{\alpha}\vert \bm{\hat{F}}\vert \Psi_{\alpha}\rangle^T \bm{R}\bm{\hat{F}}^{\rm e} \label{V_pe} .    
\end{equation}
We can write Eq.~\eqref{V_pe} in terms of individual contributions as 
\begin{align}
 \hat{V}^{\rm pe} 
                  & = \hat{V}^{\rm es} - \langle\Psi_{\alpha}\vert \bm{\hat{F}}^{\rm e} + \bm{F}^{\rm n} +\bm{F}^{\rm es}   \vert \Psi_{\alpha}\rangle^T \bm{R}\bm{\hat{F}}^{\rm e} \notag \\
                  & = \hat{V}^{\rm es} - \sum_{s' \in P}\left(\langle\Psi_{\alpha}\vert \bm{\hat{\mu}}^{\rm e}_{\text{ind},s'}\vert\Psi_{\alpha}\rangle 
+ \bm{\mu}^{\rm n}_{\text{ind},s'} +\bm{\mu}^{\rm es}_{\text{ind},s'}\right)   \bm{\hat{F}}^{\rm e}_{s'} , 
\end{align}
where $\hat{V}^{\rm pe}$ depends on $\bm{D}^{\alpha}$ in Eq.~\eqref{one-and-two-electron-dens-1} through  
\begin{align}
 \bm{\mu}^{\rm e}_{\text{ind},s'}[\bm{D}^{\alpha}] \equiv \langle \Psi_{\alpha}\vert \bm{\hat{\mu}}^{\rm e}_{\text{ind},s'}\vert \Psi_{\alpha}\rangle 
= \sum_{s''}\sum_{rs} \bm{R}_{s'',s'} \bm{F}^{\rm e}_{pq,s'} D^{\alpha}_{rs}. \label{mu_dens_dep}
\end{align}
The expectation value of $\hat{V}^{\rm pe}$ now reads  
\begin{align}
 \mathcal{E}^{\rm pe}_{\alpha}  = &\langle\Psi_{\alpha}\vert\hat{V}^{\rm pe}\vert \Psi_{\alpha}\rangle 
\equiv  \mathcal{E}^{\rm es}_{\alpha} + \mathcal{E}^{\rm pol}_{\text{es},\alpha} 
 +  \mathcal{E}^{\text{pol}}_{\text{n},\alpha}  +  \mathcal{E}^{\rm pol}_{\text{e},\alpha}[\bm{D}^{\alpha}]   \notag \\
 =& \sum_{s'}\left(E^{\rm es}_{\text{n},s'} + \sum_{pq}V^{\rm es}_{pq,s'}D^{\alpha}_{pq}\right) \notag \\
 & -  \sum_{s'}\sum_{pq} \bigl(\bm{\mu}^{\rm es}_{\text{ind},s'}\bigr)^T \bm{F}^{\rm e}_{pq}D^{\alpha}_{pq} 
 - \sum_{s'}\sum_{pq} \bigl(\bm{\mu}^{\rm n}_{\text{ind},s'}\bigr)^T \bm{F}^{\rm e}_{pq,s'}D^{\alpha}_{pq} \notag \\
 & -  \sum_{s'}\sum_{pq} \bigl(\bm{\mu}^{\rm e}_{\text{ind},s'}[\bm{D}^{\alpha}]\bigr)^T \bm{F}^{\rm e}_{pq,s'}D^{\alpha}_{pq} . \label{pe_exp_val}
\end{align}
The energy expression in Eq.~\eqref{pe_exp_val} is not linear in the 1-RDMs, as can be directly seen by inserting 
the expression for $\bm{\mu}^{\text{ind}}_{\text{e},s'}[\bm{D}^{\alpha}]$ into the last term in Eq.~\eqref{pe_exp_val}
\begin{equation}
  \mathcal{E}^{\text{pol}}_{\text{e},\alpha}[\bm{D}^{\alpha}] = - \sum_{s'}\bigl(\bm{\mu}^{\rm e}_{\text{ind},s'}[\bm{D}^{\alpha}]\bigr)^T \bm{F}^{\rm e}_{pq,s'}D^{\alpha}_{pq}  = 
- \sum_{s's''}\sum_{pqrs}(\bm{F}^{\rm e}_{rs,s'} )^T \bm{R}_{s''s'} \bm{F}^{\rm e}_{pq,s'}  D^{\alpha}_{rs} D^{\alpha}_{pq} . \label{e_pol_exp_val} 
\end{equation}
However, the optimization of a standard CAS-CI wave function assumes a linear relationship of the energy on the 1-RDMs. 
We have previously applied a work-around to this problem when adding a (non-linear) short-range DFT potential\cite{hedegaard2015b} to CAS-CI
or DMRG-CI following Pedersen\cite{pedersenphd2004}.
This will be elaborated in Section \ref{implementation_details}. As we will distinguish between 1-RDMs for the inactive and active parts,
we introduce the notation 
\begin{align}
\mathcal{E}^{\text{pol}}_{\text{e},X,\alpha}[\bm{D}^{Y,\alpha}] 
=  - \sum_{s'}\bigl(\bm{\mu}^{\rm e}_{\text{ind},s'}[\bm{D}^{Y,\alpha}]\bigr)^T \bm{F}^{\rm e}_{pq,s'}D^{X,\alpha}_{pq} , \label{e_pol_master}
\end{align}
where $X$ and $Y$ indicate the origins ('$I$' or '$A$') of the 1-RDMs.

\subsection{Accommodation of Short-Range Density Functionals} \label{CAS_srDFT}

PE-DMRG with a range-separated Hamiltonian does not create any additional coupling of the short-range DFT functional and the PE operators. The 
only difference between PE-DMRG and PE-DMRG--srDFT is their different expressions for $\hat{H}^{\text{iso}}$. To extend PE-DMRG to PE-DMRG--srDFT it  
is therefore sufficient to implement  Eq.~\eqref{Hamilton_CAS_iso} in a range-separated form\cite{fromager2007,fromager2009}  
\begin{align}
  \hat{H}^{\rm iso} \rightarrow \hat{H}^{\rm iso}_{\rm srDFT} =  \hat{H}^{\rm iso,lr} + \hat{V}^{\rm sr}_{\rm Hxc}[\rho] , 
\label{H_iso_srDFT}
\end{align}
$\hat{H}^{\rm iso,lr}$ is the long-range Hamiltonian which in this work is evaluated as an MPO (cf.~Eq.~\eqref{mpo_2}) with  
 long-range two-electron integrals separated on the basis 
of the error function (see, e.g., Eqs.~(14) and (15) in Ref.~\citenum{hedegaard2015b}). 
The contribution from DFT is introduced through the short-range Hartree--exchange--correlation 
potential, $\hat{V}^{\rm sr}_{\rm Hxc}[\rho]$,  
and the pseudo-energy corresponding to Eq.~\eqref{pseudo-energy} then reads
\begin{align}
 \mathcal{E}^{\rm srDFT}_{\alpha} = \langle\Psi^{\rm lr}_{\alpha}\vert \hat{H}^{\rm iso}_{\rm srDFT}  + \hat{V}^{\text{pe}} \vert\Psi^{\rm lr}_{\alpha}\rangle,
\label{pseudo_srdft_energy} 
\end{align}
where the integrand is the  effective PE-DMRG--srDFT Hamiltonian. 
The (long-range) DMRG wave function that diagonalizes $ \hat{H}^{\rm iso,lr}$ is 
denoted $\vert\Psi^{\rm lr}_{\alpha}\rangle$.

\section{Implementation}  \label{implementation_details}

The implementation details described in this section will mainly be concerned with the coupling of the PE model with DMRG; the implementation of the additional terms 
 for PE-DMRG--srDFT only concerns terms due to the introduction of the range-separated Hamiltonian in Eq.~\eqref{pseudo_srdft_energy}\cite{hedegaard2015b}, 
which amount to the evaluation of the short-range Hartree--exchange--correlation potential, $\hat{V}^{\rm sr}_{\rm Hxc}[\rho]$, 
in $\langle\Psi^{\rm lr}_{\alpha}\vert \hat{H}^{\rm iso}_{\rm srDFT}\vert \Psi^{\rm lr}_{\alpha} \rangle$. A full account of the evaluation of this term was given  
for DMRG--srDFT in Ref.~\citenum{hedegaard2015b}. We will in fact employ the same technique as in Ref.~\citenum{hedegaard2015b} to evaluate the 
non-linar part of the PE operator and we therefore introduce the deviation, $\Delta D^{\alpha}_{pq}$, 
\begin{align}
 \Delta D^{\alpha}_{pq} =  D^{\alpha}_{pq} - D^{\text{ref},\alpha}_{pq},
\label{delta_D_pq_1} 
\end{align}
from some (fixed) reference density matrix, $\bm{D}^{\text{ref},\alpha} = \{ D^{\text{ref},\alpha}_{pq}\}$, 
\begin{align}
D^{\text{ref},\alpha}_{pq} = D^{I,\alpha}_{ij} + D^{\text{ref},A,\alpha}_{uv},  
\label{fixation} 
\end{align}
so that the active-part 1-RDM elements can be written as
\begin{equation} 
D^{\text{A},\alpha}_{uv} = D^{\text{ref,A},\alpha}_{uv} + \Delta D^{\text{A},\alpha}_{uv}  . \label{delta_D_pq_2} 
\end{equation}
Note that the linearization in Eq.~\eqref{delta_D_pq_1} is an approximation only required to obtain a self-consistent scheme. 
The final (converged) state does not involve other approximations than those inherent to PE, DMRG, and srDFT.
As we seek to divide contributions into active and inactive parts, as in Eqs.~\eqref{inact-energy} and \eqref{act-energy}, 
\begin{equation}
 \mathcal{E}^{\rm pe}_{\alpha} =  \mathcal{E}^{\rm pe}_{I,\alpha} +  \mathcal{E}^{\rm pe}_{A,\alpha}  ,
\end{equation}
we obtain with Eq.~\eqref{delta_D_pq_1} for the non-linear part in Eq.~\eqref{pe_exp_val} 
\begin{align}
  \mathcal{E}^{\rm pe}_{I,\alpha} & = \mathcal{E}^{\rm es}_{I,\alpha} + \mathcal{E}^{\rm pol}_{\text{es},I,\alpha} + \mathcal{E}^{\rm pol}_{\text{n},I,\alpha}  + 
\mathcal{E}^{\rm pol}_{\text{e},\text{ref},\alpha}[\bm{D}^{\text{ref},\alpha}] ,   
\label{E_pe_I_1} 
\\
  \mathcal{E}^{\rm pe}_{A,\alpha} & = \mathcal{E}^{\rm es}_{A,\alpha} + \mathcal{E}^{\rm pol}_{\text{es},A,\alpha} + \mathcal{E}^{\rm pol}_{\text{n},A,\alpha} 
+ 2\Delta \mathcal{E}^{\rm pol}_{\text{e},A,\alpha}[\bm{D}^{\text{ref},A,\alpha}]  . 
\label{E_pe_A_1}
\end{align}
The pseudo-energies $\mathcal{E}^{\rm es}_{A,\alpha}$, $\mathcal{E}^{\rm pol}_{\text{es},A,\alpha}$, $\mathcal{E}^{\rm pol}_{\text{n},A,\alpha}$ 
and the corresponding inactive parts are given as in Eq.~\eqref{pe_exp_val}, replacing  
$D^{\alpha}_{pq}\rightarrow D^{A,\alpha}_{uv}$ and  $D^{\alpha}_{pq}\rightarrow D^{I,\alpha}_{ij}$. The terms 
 $\mathcal{E}^{\rm pol}_{\text{e},\text{ref},\alpha}[\bm{D}^{\text{ref},\alpha}]$ and $\Delta\mathcal{E}^{\rm pol,\text{ref}}_{\text{e},A,\alpha}[\bm{D}^{\text{ref},A,\alpha}]$ 
associated with inactive and active parts, respectively, are defined 
in accordance with Eq.~\eqref{e_pol_master} and therefore given as 
\begin{equation}
  \mathcal{E}^{\text{pol},\alpha}_{\text{e},\text{ref},\alpha}[\bm{D}^{\text{ref},\alpha}] 
=  -\sum_{s'\in P}\sum_{pq} \bigl(\bm{\mu}^{\rm ind}_{\text{e},s'}[\bm{D}^{\text{ref},\alpha}]\bigr)^T \bm{F}^{\rm e}_{pq,s'}D^{\text{ref},\alpha}_{pq} 
\label{E_pol_e_I} 
\end{equation}
and 
\begin{equation}
  \Delta \mathcal{E}^{\text{pol}}_{\text{e},A,\alpha}[\bm{D}^{\text{ref},A,\alpha}] = 
-\sum_{s'\in P}\sum_{uv} \bigl(\bm{\mu}^{\rm ind}_{\text{e},s'}[\bm{D}^{\text{ref}, A,\alpha}]\bigr)^T \bm{F}^{\rm e}_{uv,s'} \Delta D^{A,\alpha}_{uv} .  
\label{E_pol_e_A_1} 
\end{equation} 
With Eq.~\eqref{delta_D_pq_2} and the reference density matrix being fixed by Eq.\ (\ref{fixation}),
we can rewrite Eqs.~\eqref{E_pe_I_1} and \eqref{E_pe_A_1} to become
\begin{align}
  \mathcal{E}^{\rm pe}_{I,\alpha} &= \mathcal{E}^{\rm es}_{I,\alpha} + \mathcal{E }^{\rm pol}_{\text{es},I,\alpha} + \mathcal{E}^{\rm pol}_{\text{n},I,\alpha} 
+ \mathcal{E}^{\rm pol}_{\text{e},I,\alpha}[\bm{D}^{I,\alpha}] -    \mathcal{E}^{\rm pol}_{\text{e},\text{ref},A,\alpha}[\bm{D}^{\text{ref},A,\alpha}] , \label{E_pe_I} 
\end{align}
and 
\begin{align}
   \mathcal{E}^{\rm pe}_{A,\alpha} &= \mathcal{E}^{\rm es}_{A,\alpha} + \mathcal{E}^{\rm pol}_{\text{es},A,\alpha} + \mathcal{E}^{\rm pol}_{\text{n},A,\alpha} 
+ 2\bigl(\mathcal{E}^{\rm pol}_{\text{e},A,\alpha}[\bm{D}^{I,\alpha}] + \mathcal{E}^{\rm pol}_{\text{e},A,\alpha}[\bm{D}^{\text{ref},A,\alpha}] \bigr),  \label{E_pe_A}
\end{align}
where the $\mathcal{E}^{\text{pol}}_{\text{e},\alpha}[\bm{D}^{\alpha}]$ terms in Eqs.~\eqref{E_pe_I} and \eqref{E_pe_A}
are defined according to the notation in Eq.~\eqref{e_pol_master}. These last two equations are the operational expressions implemented. 
For the extension to PE-DMRG--srDFT, from Eq.~\eqref{pseudo_srdft_energy} it follows that we obtain expressions that are equilvalent to 
Eqs.~\eqref{E_pe_I} and \eqref{E_pe_A} from $\langle\Psi^{\rm lr}_{\alpha}\vert \hat{V}^{\rm pe}\vert \Psi^{\rm lr}_{\alpha}\rangle$. In 
addition, we obtain a term from $\hat{H}^{\rm iso}_{\rm srDFT}$ due to the short-range Hartree--exchange--correlation potential, $\hat{V}^{\rm sr}_{\rm Hxc}[\rho]$. 
 As shown in Ref.~\citenum{hedegaard2015b}, these terms can be treated similarly to the effective PE operator in Eq.~\eqref{V_pe}, i.e.,~by 
the method of linearization  in Eq.~\eqref{delta_D_pq_1}. The resulting terms are given in Eqs.~(37) and (39) of Ref.~\citenum{hedegaard2015b}.   

The theory described above  was implemented in a development version of the {\sc Dalton} program\cite{DALTON2016}. 
This version of {\sc Dalton} includes an interface to the release version of the PE Library\cite{pelib}, {\sc pelib}, and an interface 
to our MPO-based DMRG program, {\sc QCMaquis}\cite{keller2014,dolfi2014,keller2015,knecht2016,keller2016}.

\section{Computational methodology}\label{compmet}

The PE  potentials constructed in this work represent a water cluster of 127 water molecules.  
The potentials were obtained with a python script that automates the extraction of coordinates for each fragment (here a water molecule) 
and sets up the calculation of localized multipoles and polarizabilities.\cite{jmothesis} 
The latter are obtained with the LoProp procedure\cite{gagliardi2004} as implemented in {\sc Molcas}\cite{aquilante2010}. 
The PE potentials include multipoles up to quadrupoles and anisotropic polarizabilities, which will 
be denoted M2P2. In addition, we have constructed potentials including only charges (M0) and 
charges, dipoles, and quadrupoles (M2). The LoProp procedure requires specially constructed basis sets of which we used 
A-6-31GPG (M2P2) and A-AUG-CC-PVTZ (M2P2, M2 and M0).\cite{gagliardi2004}   
The coordinates of the water cluster were taken from the supporting information of Ref.~\citenum{jacob2006} and 
have already served in a range of previous works\cite{jacob2006,jensen2003a,jensen2003b,kongsted2002}. Therefore, they represent an excellent benchmark for our approach.
The structure for the water cluster can be seen as a representative structure for an ensemble of structures. 
It was originally obtained in Ref.~\citenum{kongsted2002} 
by averaging 8000 molecular dynamics (MD) trajectories into one effective structure,
where each trajectory describes a box of 128 water molecules. The MD was performed\cite{kongsted2002} with 
a polarizable force field and periodic boundary conditions in which the individual trajectories spanned 20 picoseconds, starting from different velocities.

\begin{figure}[tbh!]
\centering
 \includegraphics[scale=0.40]{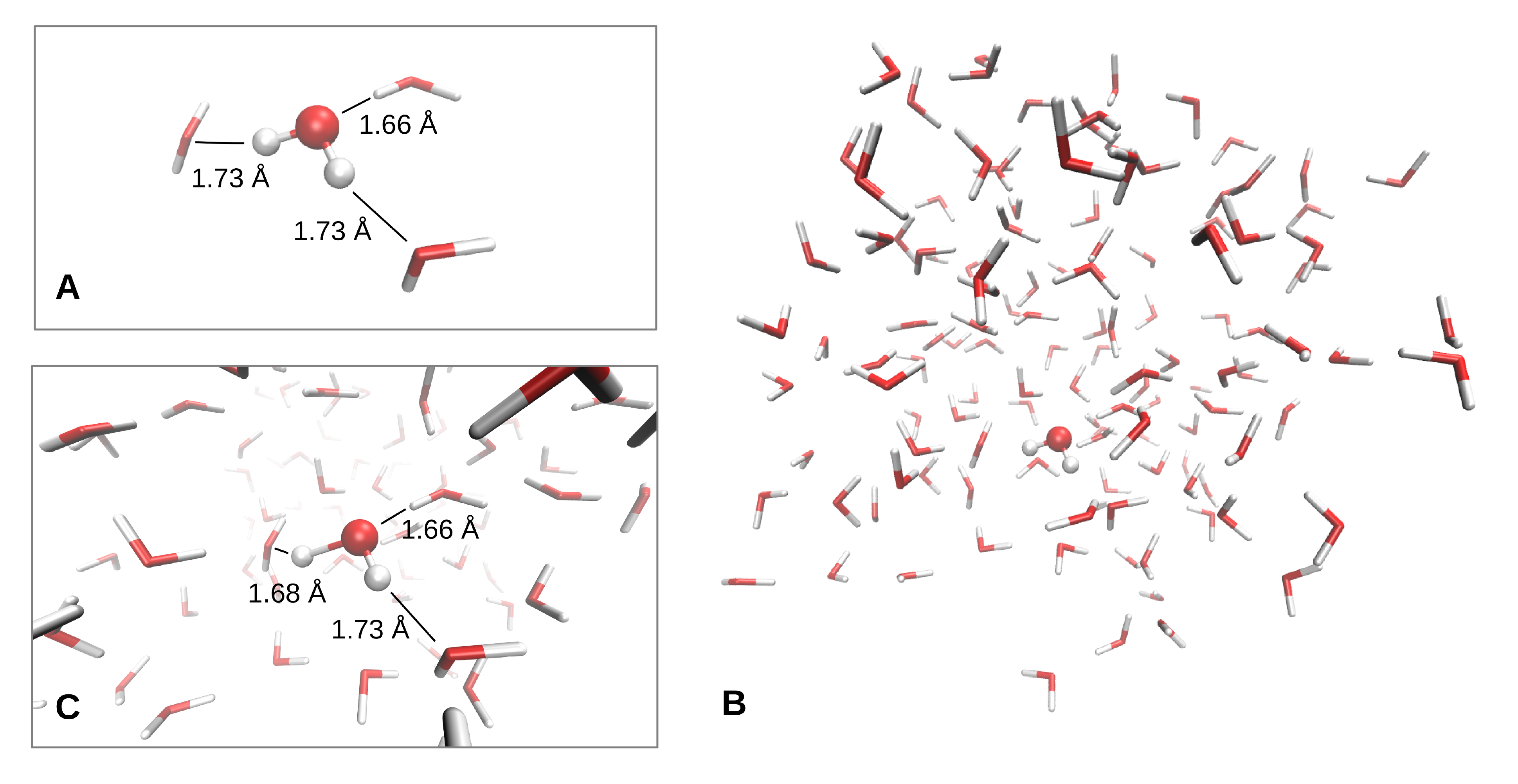}
 \caption{Water cluster from Ref.~\citenum{kongsted2002}. The QM region is shown in ball-and-stick representation. 
(A) Cut-out of the single water molecule taken as the QM region including three nearest-neighbor water molecules. 
(B) Full structure. (C) Zoom in on the QM water molecule where one of the bonds was elongated by a factor of 3/2.  
\label{christoph_h2o_full}}
\end{figure}

The QM water molecule and a minimal solvent shell with only three nearest neighbors involved in hydrogen bonding 
are shown in Figure \ref{christoph_h2o_full}A, whereas the full 
cluster is shown in Figure \ref{christoph_h2o_full}B. We also constructed a system that exhibits multiconfigurational character to a certain degree by elongating 
one of the water {H--O} bonds as shown in Figure \ref{christoph_h2o_full}C.  

As an example of a QM subsystem requiring a large active space in a structured environment, we chose the
the retinal chromophore in rhodopsin. This system has become a benchmark system for advanced electronic structure methods
in recent years (see, for instance, Refs.\ \citenum{ferre2003,andruniow2004,cembran2005,roth05,coto06,roth07,frut07,stra08,scha11,wein11,roth13,roth14,roth14b,vals15,luk-15}).
For the channelrhodopsin protein, we employ the M2P2 potential constructed in 
Ref.~\citenum{hedegaard2015a}, which assumes a QM region comprised of the retinalidyne chromophore and a small part of the lysine (lys296) side chain  
that links the chromophore to the protein (depicted below in Figure \ref{mut_inf_gs_dmrg-20-27_m2p2}, middle).  

All DMRG calculations for the QM parts were carried out with {\sc QCMaquis}\cite{keller2014,dolfi2014,keller2015,knecht2016,keller2016} 
in a Hartree--Fock (HF) or HF--srDFT molecular orbital basis. The corresponding PE-DMRG and PE-DMRG--srDFT calculations employed 
PE-HF and PE-HF--srDFT molecular orbital bases, respectively. For water, the DMRG calculations employed 
512 renormalized block states $m$, whereas for the retinalidyne chromophore this number was increased to 1024 renormalized block states.  
The importance of diffuse functions in the first excitation of the water molecule was previously highlighted\cite{jacob2006}. We chose a 
6-31++G* basis set\cite{hehre1972,harihara1973,clark1983}, meaning that DMRG(10,30)[512] corresponds 
to the active space of a full configuration interaction (FCI) calculation.    
However, although this small basis set will not yield accurate results, it is a suitable choice here as it
permits us to carry out a FCI-type DMRG calculation that does not require further consideration of dynamical correlation effects.

For the retinylidene chromophore, we investigated the multiconfigurational nature of the ground and first excited states in terms of 
single-orbital entropies\cite{legeza2003a} and mutual information\cite{legeza2006,rissler2006} evaluated for the frontier orbitals. 
For this analysis, the DMRG(20,27)[1024] active space encompassed 
the $\pi$-system and higher lying $\sigma$/$\sigma^{*}$ orbitals. The same active space was employed in the DMRG(20,27)[1024]--srDFT 
calculations. The latter calculations 
were performed with the tailored short-range PBE functional by Goll, Werner, and Stoll\cite{goll2005}, and we denote these
calculations DMRG(20,27)[1024]--srPBE in the following.      
All calculations on retinylidene employed a 6-31G* basis set because this very small basis had already been chosen for 
studies on related rhodopsins\cite{andruniow2004,altun2009} and hence makes a comparison possible. 
Despite the fact that this basis set is very small, previous calculations on channelrhodopsin and a related rhodopsin showed that 
the $S_0 \rightarrow S_1$ excitation is not too sensitive 
to an increasing basis set size.\cite{altun2008,sneskov2013} 

For the water systems we have in addition to the PE-DMRG calculations, also carried out PE-TDDFT calculations with 
four different density functionals, namely BLYP\cite{becke1988,lee1988}, B3LYP\cite{becke1988,lee1988,becke1993}, PBE\cite{perdew1996b}, 
and PBE0\cite{adamo1999}. Due to the Rydberg character of the excitation, we investigated the effect of a range-separated functional,
for which we chose the long-range corrected CAM-B3LYP functional.\cite{yanai2004} The effect is, in this case, about 0.2 eV compared to B3LYP and hence comparatively small.
Although solvent shifts and absolute excitation energies differ between the functionals, the obtained trends between different PE potentials are similar. 
We therefore chose  B3LYP as representative for the DFT results and only detailed B3LYP data are given explicitly; 
results obtained with the other functionals are only shown in the figures.

\section{Results and discussion}\label{results}

Within a state-specific approach, we investigated the effect of the environment on the first electronic excitation ($1^{1}A_1 \rightarrow 1^{1}B_1 $) of a water molecule. 
The experimental value for this excitation  is known in both gas and condensed phase. In the gas phase, the excitation energy 
is approximately 7.4 eV whereas it is blue-shifted to about 8.2 eV in condensed phase, yielding a solvent shift of about 0.8 eV. A discussion 
of the experimental results can be found in the paper by Christiansen et al.\cite{christiansen2000}.  
In the first subsection (Section \ref{small_environment}), we consider the small solvent shell of three water molecules as the environment 
that is to be represented by a PE potential (Figure \ref{christoph_h2o_full}A). In the second 
subsection (Section \ref{large_environment}), we extend the environment so that all 127 water molecules are represented by a PE potential (Figure \ref{christoph_h2o_full}B). 
To highlight the differences between our PE-DMRG approach and PE-TDDFT for multiconfigurational cases, 
we consider in Section \ref{stretched_OH_bond} the solute with the elongated {O--H} bond (Figure \ref{christoph_h2o_full}C) 
where the multireference character is incorporated already in the electronic ground state.   

Although we also compare with experimental results, 
it should be emphasized that results obtained with only one structural configuration must be considered with care. For an accurate comparison, 
one must employ a stochastically meaningful number of snapshots. However, our main purpose is to compare the various embedding potentials 
for state-specific approaches within the PE approach.

Afterwards we consider the retinylidene chromophore within its host channelrhodopsin protein (Section \ref{channelrhodopsin}). 
For this system, the protein environment is known to cause a large blue-shift of 
the first excitation energy\cite{sneskov2013}. DMRG studies of chromophores  within proteins are rare 
(see, e.g., Ref.~\citenum{sharma2014b} 
where a point-charge description of the environment was invoked for a iron-sulfur cluster).

\subsection{Water Molecule Embedded in a Minimal Solvent Shell}\label{small_environment}

The state-specific excitation energies obtained with PE-DMRG and the predicted solvent 
shifts are given in Table \ref{h2o-pe-dmrg-2} (TDDFT an PE-TDDFT results with B3LYP are also given in this table).   
\begin{figure}[tbh!]
\centering
  \includegraphics[scale=0.43]{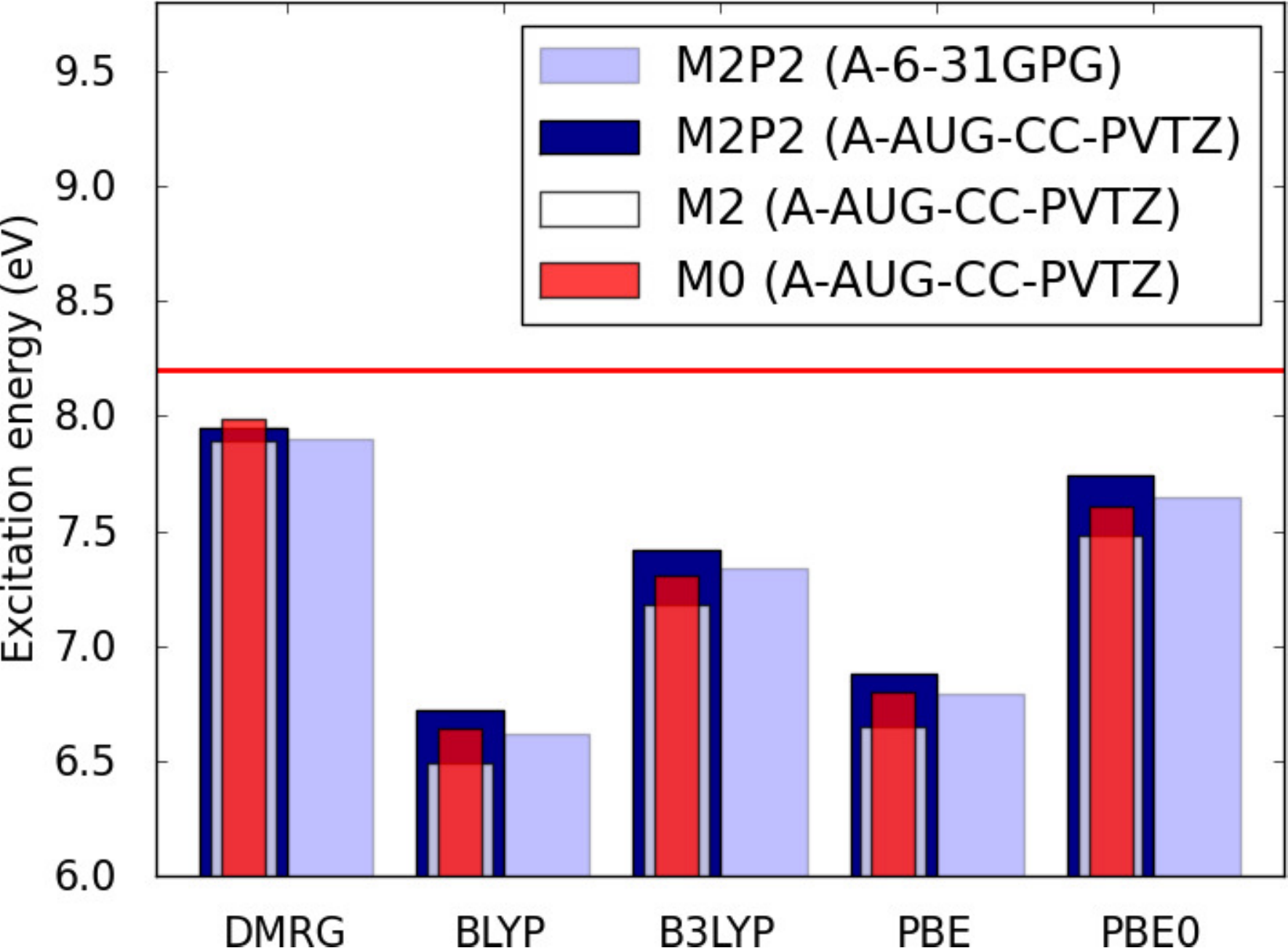}
   \hspace{0.5cm}
 \includegraphics[scale=0.43]{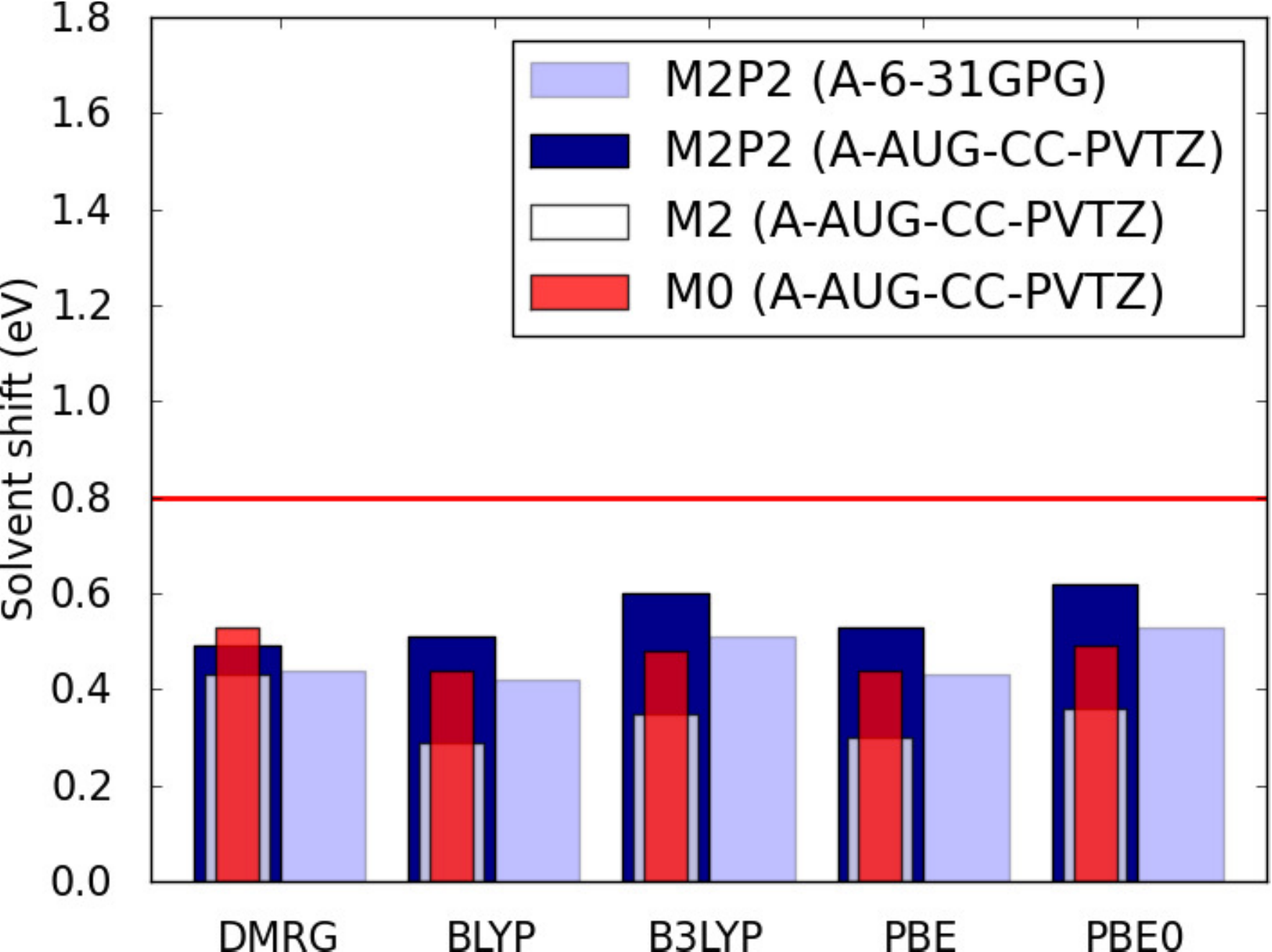}
 \caption{Excitation energies and solvent shifts for the environment comprised of three water molecules, see Figure \ref{christoph_h2o_full}A. 
The column for the A-AUG-CC-PVTZ basis set collects contributions from charges (M0), charges, dipoles and quadrupoles (M2), and 
charges, dipoles, quadrupoles, and polarizabilities (M2P2).
The experimental result is shown as a red, horizontal line.  \label{trimer}}
\end{figure}
\begin{table}
\caption{Excitation energies and solvent shifts (in eV) for {H$_2$O} in PE potentials representing three environment water molecules.  
 \label{h2o-pe-dmrg-2}} 
 \begin{tabular}{lcccc}
\hline \hline \\[-1.5ex]
%--------------------------------------------------------------------------------------
Potential & \multicolumn{2}{c}{Exc. Energy}  & \multicolumn{2}{c}{Solvent shift}  \\
\hline \\[-1.0ex ]
%--------------------------------------------------------------------------------------
                        & ~~DMRG(10,30)[512]~~ & ~~B3LYP~~ &  ~~DMRG(10,30)[512]~~ & ~~B3LYP ~~ \\[0.5ex] 
Isolated                &  7.46                &  6.83     &        -              &     -   \\[0.5ex] 
M0$^a$                  &  7.99                &  7.31     &      0.53             &  0.48      \\[0.5ex]
M2$^a$                  &  7.89                &  7.18     &      0.43             &  0.35      \\[0.5ex]
M2P2$^b$                &  7.90                &  7.34     &      0.44             &  0.51      \\[0.5ex]
M2P2$^a$                &  7.95                &  7.42     &      0.49             &  0.59      \\[0.5ex]
%--------------------------------------------------------------------------------------
\hline \hline
 \end{tabular} \\
\noindent {\footnotesize $^a$Obtained with A-AUG-PTVZ/B3LYP} \\
\noindent {\footnotesize $^b$Obtained with A-6-31PGP/B3LYP }
\end{table}
Starting with the result for the isolated {H$_2$O} molecule, we find that the DMRG(10,30)[512] excitation energy 
is in good agreement with the experimental value of 7.4 eV. 
 As one would have expect from the significant Rydberg character of the excitation\cite{jacob2006},  
the excitation energies from TDDFT and PE-TDDFT are underestimated.  
We note that this can be partially corrected by  asymptotically corrected functionals, although we find that the effect is rather small (see Section \ref{compmet}). 
In fact, the result for the isolated {H$_2$O} molecule from Ref.~\citenum{jacob2006} 
(7.76 eV) is slightly overestimated with the SAOP potential\cite{schipper2000} in combination with a Slater-type basis set including diffuse functions.  
  
For the embedding calculations, the state-specific calculation with the most accurate potential yields
an excitation energy of 7.95 eV and thereby a solvent shift of 0.49 eV (see Table \ref{h2o-pe-dmrg-2}). 
The experimental solvent shift is therefore underestimated by about a factor two. 
By comparing the M0 entry with the M2 and M2P2 entries in Table \ref{h2o-pe-dmrg-2}, we see that M0 alone produces a solvent shift of 0.44 eV and hence 
the effect from the charges is by far the largest. The additional 
effect from higher-order multipoles (M2) amounts to 0.10 eV and is in the opposite direction, thereby reducing the total shift slightly. 
Increasing the accuracy by including polarizabilities has little effect for this small environment, but (as will be 
apparent in the next subsection) becomes nonnegligible when considering a larger environment. 

Regarding the PE-TDDFT results, the absolute excitation energies and solvent shifts of 
all density functionals are compared to the PE-DMRG(10,30)[512] result in Figure \ref{trimer}.  
The excitation energies and solvent shifts show the same trends 
 as the state-specific PE-DMRG 
results with respect to the applied potential. Yet, the PE-TDDFT excitation energies and 
solvent shifts show a larger variation with respect to how the environment potentials are obtained than our PE-DMRG results. 
 As expected, the absolute excitation energies are too low and the solvent shifts are predicted lower than the experimental one. 
The latter was also observed in Ref.~\citenum{jacob2006} for a small environment comprising only two water molecules. 

\subsection{Water Molecule Embedded in Larger Environment}\label{large_environment}

We now consider the larger environment of 127 water molecules (Figure \ref{christoph_h2o_full}B). 
The excitation energies and solvent shifts for PE-DMRG and PE-TDDFT (with B3LYP) are given in Table \ref{h2o-pe-dmrg-full}. 
\begin{table}
\caption{Excitation energies and solvent shifts (in eV) for a solvated system of {H$_2$O} 
in PE potentials representing 127 environment water molecules.  
 \label{h2o-pe-dmrg-full}} 
 \begin{tabular}{lcccc}
\hline \hline \\[-1.5ex]
%--------------------------------------------------------------------------------------
Potential & \multicolumn{2}{c}{Exc. Energy}  & \multicolumn{2}{c}{Solvent shift}  \\
\hline \\[-1.0ex ]
%--------------------------------------------------------------------------------------
                        & ~~DMRG(10,30)[512]~~ & ~~B3LYP~~ &  ~~DMRG(10,30)[512]~~ & ~~B3LYP ~~ \\[0.5ex] 
M0$^a$                  &      8.21            &   7.53    &         0.75          &    0.70    \\[0.5ex]
M2$^a$                  &      8.04            &   7.32    &         0.58          &    0.49    \\[0.5ex]
M2P2$^b$                &      8.17            &   7.65    &         0.71          &    0.82    \\[0.5ex]
M2P2$^a$                &      8.36            &   7.89    &         0.90          &    1.06    \\[0.5ex]
%--------------------------------------------------------------------------------------
\hline \hline
 \end{tabular} \\
\noindent {\footnotesize $^a$Obtained with A-AUG-PTVZ/B3LYP} \\
\noindent {\footnotesize $^b$Obtained with A-6-31PGP/B3LYP }
\end{table}

\begin{figure}[tbh!]
\centering
  \includegraphics[scale=0.43]{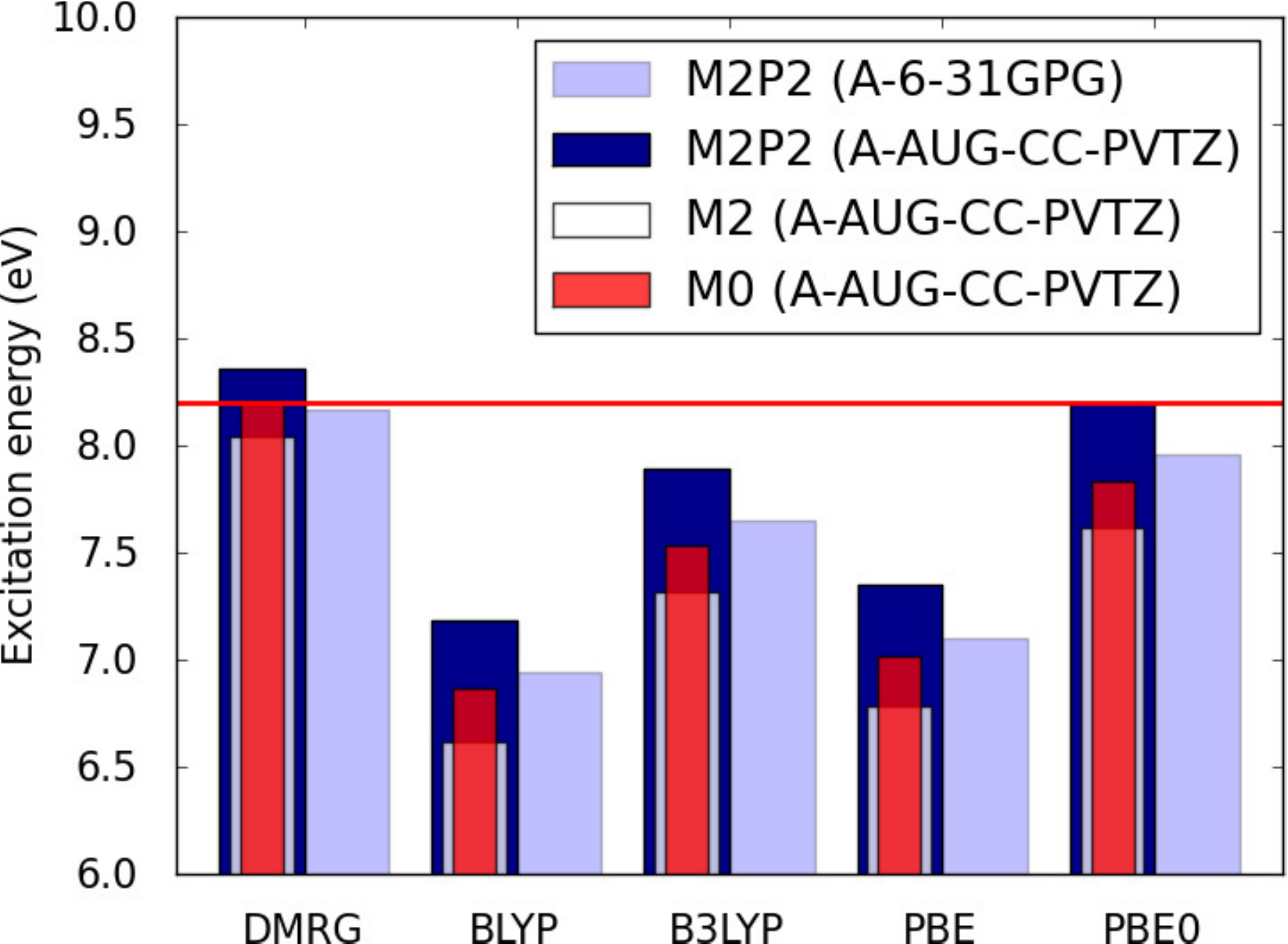}
   \hspace{0.5cm}
 \includegraphics[scale=0.43]{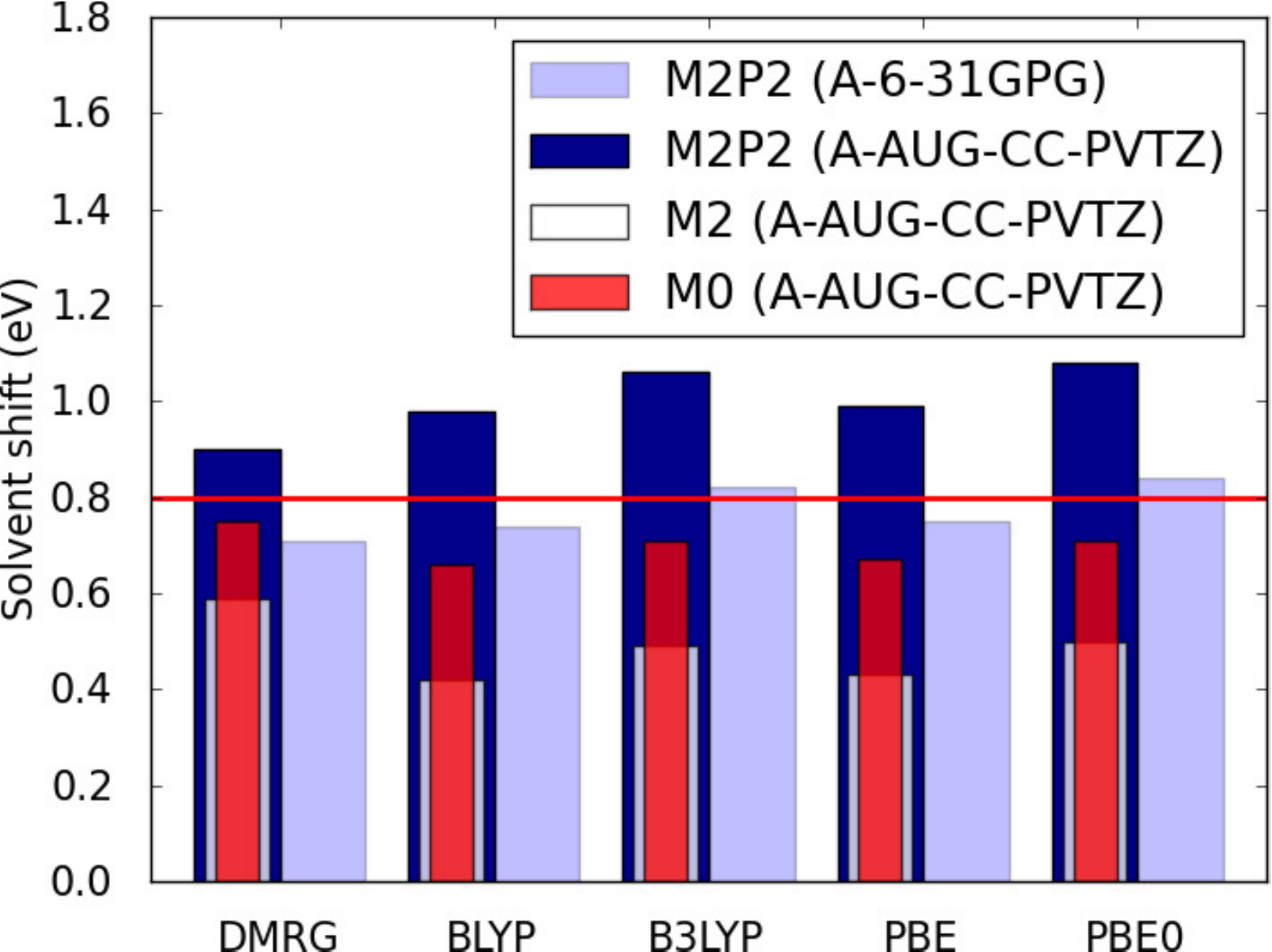}
 \caption{Excitation energies and solvent shifts for the full water cluster in Figure \ref{christoph_h2o_full}B. 
The column for the A-AUG-CC-PVTZ basis set collects contributions from charges (M0), charges, dipoles and quadrupoles (M2), and 
charges, dipoles, quadrupoles and polarizabilities (M2P2). 
The experimental result is shown as a red, horizontal line.  \label{full_system}}
\end{figure}

Compared to the small system described in Section \ref{small_environment}, the excitation energies and solvent shifts increase. 
For PE-DMRG with the large A-AUG-PVTZ basis set, 
the charges alone give an excitation energy that is almost on top of the experimental value, which must be considered fortuitous. 
From Table \ref{h2o-pe-dmrg-full} (see also Figure \ref{full_system}) we see that 
including higher-order multipoles lowers the excitation energy (by 0.17 eV), while polarization again counteracts by raising the excitation energy (by 0.32 eV). 
Hence, our most accurate DMRG result overestimates the excitation energy compared to experiment, albeit by a small margin.  
Compared to the small environment (Section \ref{small_environment}), the effect of polarization gains importance.  
Considering that the DMRG excitation energies obtained with the two potentials including polarization (M2P2) differ by 0.19 eV shows 
that accurate potentials require fairly large basis sets.
This was also concluded in a DFT study\cite{olsen2015b} of amino acids in their ground state. 
In this regard, it must be considered fortuitous, as for the case with M0 charges above,
that the DMRG(10,30)[512] result with the M2P2 potential 
constructed with A-6-31GPG/B3LYP is very close to the experimental value.  
As can be seen from Table \ref{h2o-pe-dmrg-full} and Figure \ref{full_system}, PE-TDDFT overestimates the effect of 
polarization compared to PE-DMRG (and also with respect to the experimental result), 
leading to an overall overestimation of the solvent shift. PE-DMRG also overestimates the experimental shift, but the overestimation is reduced, compared to PE-TDDFT. 
Such an overestimation was also found within a FDE-TDDFT scheme for this particular system\cite{jacob2006}.
With the exception of the 
PBE0 functional, the DFT results underestimate the absolute excitation energies compared to PE-DMRG (and also compared to the experimental result). 
The large scatter in the excitation energies and solvent shifts from the different density functionals point to the value of 
DFT-independent data provided by wavefunction methods (note, however, that DMRG results in general lack dynamic-correlation contributions, which
is of no importance here as our DMRG calculations are actually FCI calculations in the one-electron basis sets chosen).

\subsection{Water Molecule with Stretched O-H Bond Embedded in Larger Environment}\label{stretched_OH_bond}

\begin{figure}[tbh!]
\centering
  \includegraphics[scale=0.43]{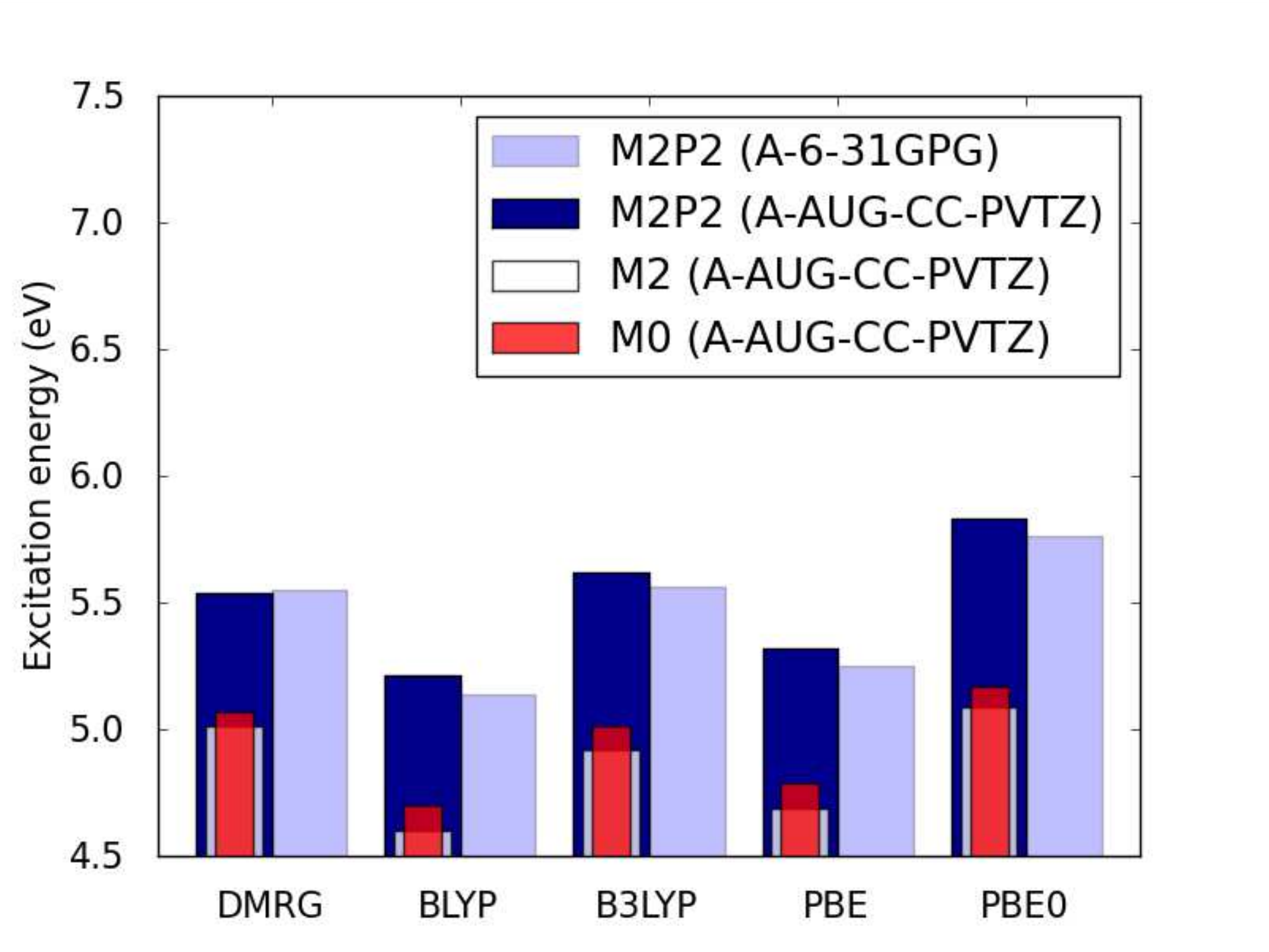}
   \hspace{0.5cm}
 \includegraphics[scale=0.43]{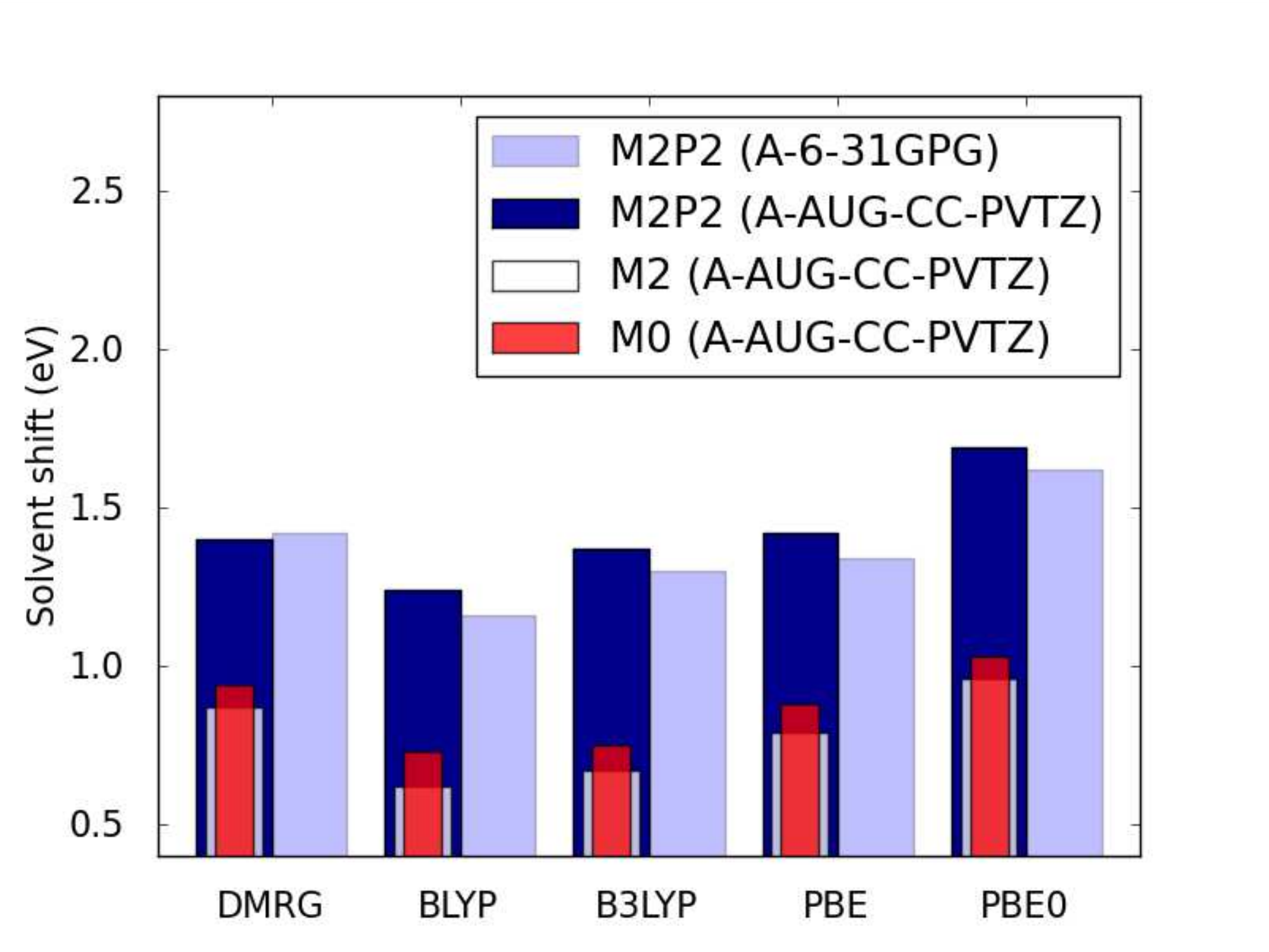}
 \caption{Excitation energies and solvent shifts for the full water cluster in Figure \ref{christoph_h2o_full}C. 
The column for the A-AUG-CC-PVTZ basis set collects contributions from charges (M0), charges, dipoles and quadrupoles (M2), and 
charges, dipoles, quadrupoles and polarizabilities (M2P2).  
 \label{full_system_1-5}}
\end{figure}

The system shown in Figure \ref{christoph_h2o_full}C was constructed by elongating one bond of the QM water molecule by a factor 1.5 (to $1.435$ {\AA}). 
Thereby, the natural occupation number for the orbital involved in the {O--H} bond decreases from 1.97 to 1.93 (for the ground state), indicating 
increased multiconfigurational character.   
The excitation energies and solvent shifts are given in Table \ref{h2o-pe-dmrg-full_1-5}. 
Both PE-DMRG(10,30)[512] and PE-TDDFT (B3LYP) results show that the excited state is destabilized more than the ground state upon bond elongation. Hence, 
the excitation energy is significantly lowered compared to the system with bond length derived from MD simulations (Table \ref{h2o-pe-dmrg-full}). 
This conclusion also holds for results obtained with the other functionals (cf.~\ref{full_system_1-5}, left).   
\begin{table}
\caption{Excitation energies and solvent shifts (in eV) for a solvated system of {H$_2$O} with stretched {O--H} bond 
in PE potentials representing 127 environment water molecules. 
 \label{h2o-pe-dmrg-full_1-5}} 
 \begin{tabular}{lcccc}
\hline \hline \\[-1.5ex]
%--------------------------------------------------------------------------------------
Potential & \multicolumn{2}{c}{Exc. Energy}  & \multicolumn{2}{c}{Solvent shift}  \\
\hline \\[-1.0ex ]
%--------------------------------------------------------------------------------------
                        & ~~DMRG(10,30)[512]~~ & ~~B3LYP~~ &  ~~DMRG(10,30)[512]~~ & ~~B3LYP ~~ \\[0.5ex] 
Isolated                  &       4.14           &   4.25    &          -            &   -        \\[0.5ex]
M0$^a$                  &       5.07           &   5.01    &          0.93         &  0.76      \\[0.5ex]
M2$^a$                  &       5.01           &   4.92    &          0.87         &  0.67      \\[0.5ex]
M2P2$^b$                &       5.55           &   5.56    &          1.41         &  1.31      \\[0.5ex]
M2P2$^a$                &       5.54           &   5.62    &          1.40         &  1.37      \\[0.5ex]
%--------------------------------------------------------------------------------------
\hline \hline
 \end{tabular} \\
\noindent {\footnotesize $^a$Obtained with A-AUG-PTVZ/B3LYP} \\
\noindent {\footnotesize $^b$Obtained with A-6-31PGP/B3LYP }
\end{table}
In particular, the influence of polarization is increased for the system with elongated {O--H} bond. Naturally, this is not only due to bond elongation, but also partly 
due to the decreased distance between the hydrogen atom from the elongated {O--H} bond and environment water molecules (see Figure \ref{christoph_h2o_full}). In such cases, it 
can have some effect to include quantum mechanical corrections in the PE scheme, as suggested in a recent paper\cite{olsen2015a}. Our current PE-DMRG method 
does not include such correction schemes, but the approach from Ref.~\citenum{olsen2015a} can be straightforwardly extended to a DMRG wave function. 
We finally note that, as expected for a system that displays multiconfigurational character, the PE-TDDFT results in Figure \ref{full_system_1-5} 
show a large spread of absolute excitation energies, with a clustering of results for pure density functionals (BLYP and PBE) and
hybrid density functionals (B3LYP and PBE0).

\subsection{Retinylidene in Channelrhodopsin \label{channelrhodopsin}}

\begin{figure}[ht] \label{ fig7} 
    \includegraphics[width=.40\linewidth]{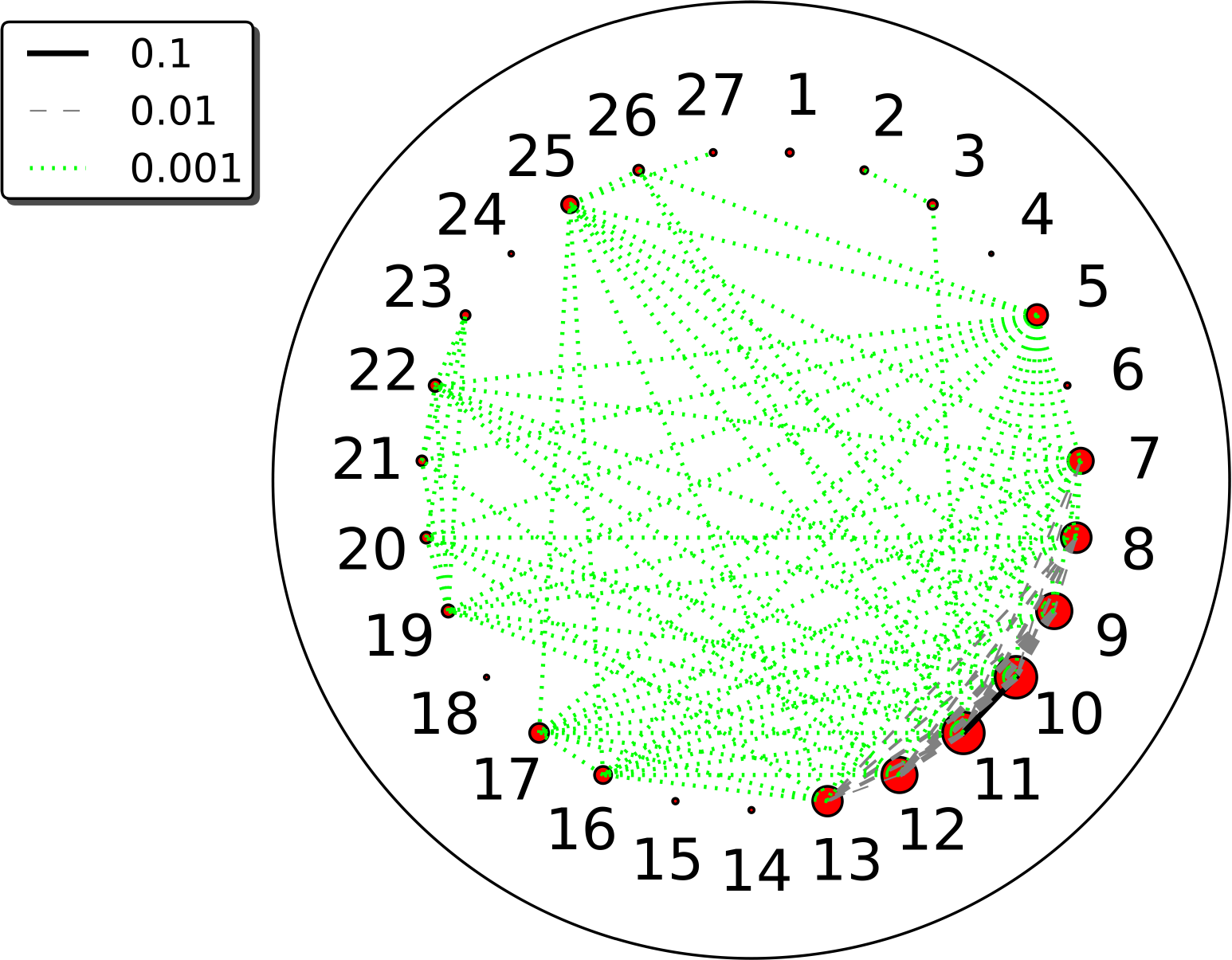}
     \hspace{0.5cm} 
        \includegraphics[width=.10\linewidth]{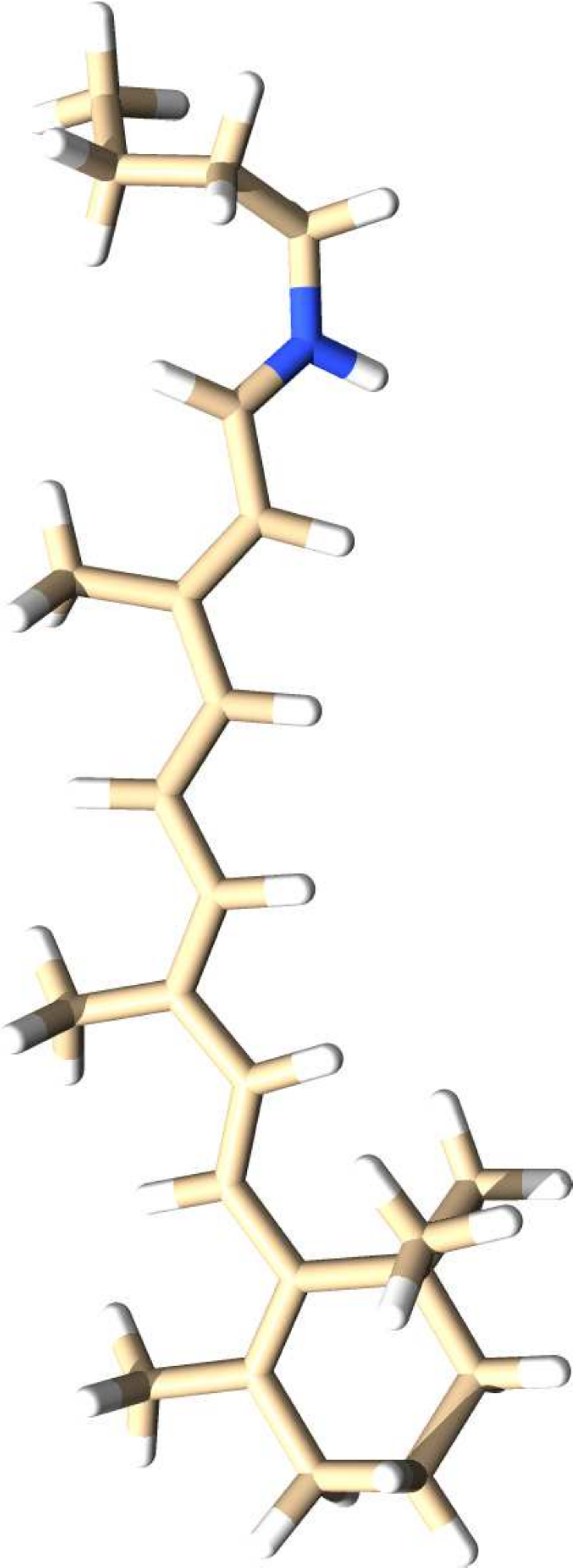} 
     %\hspace{0.5cm} 
    \includegraphics[width=.40\linewidth]{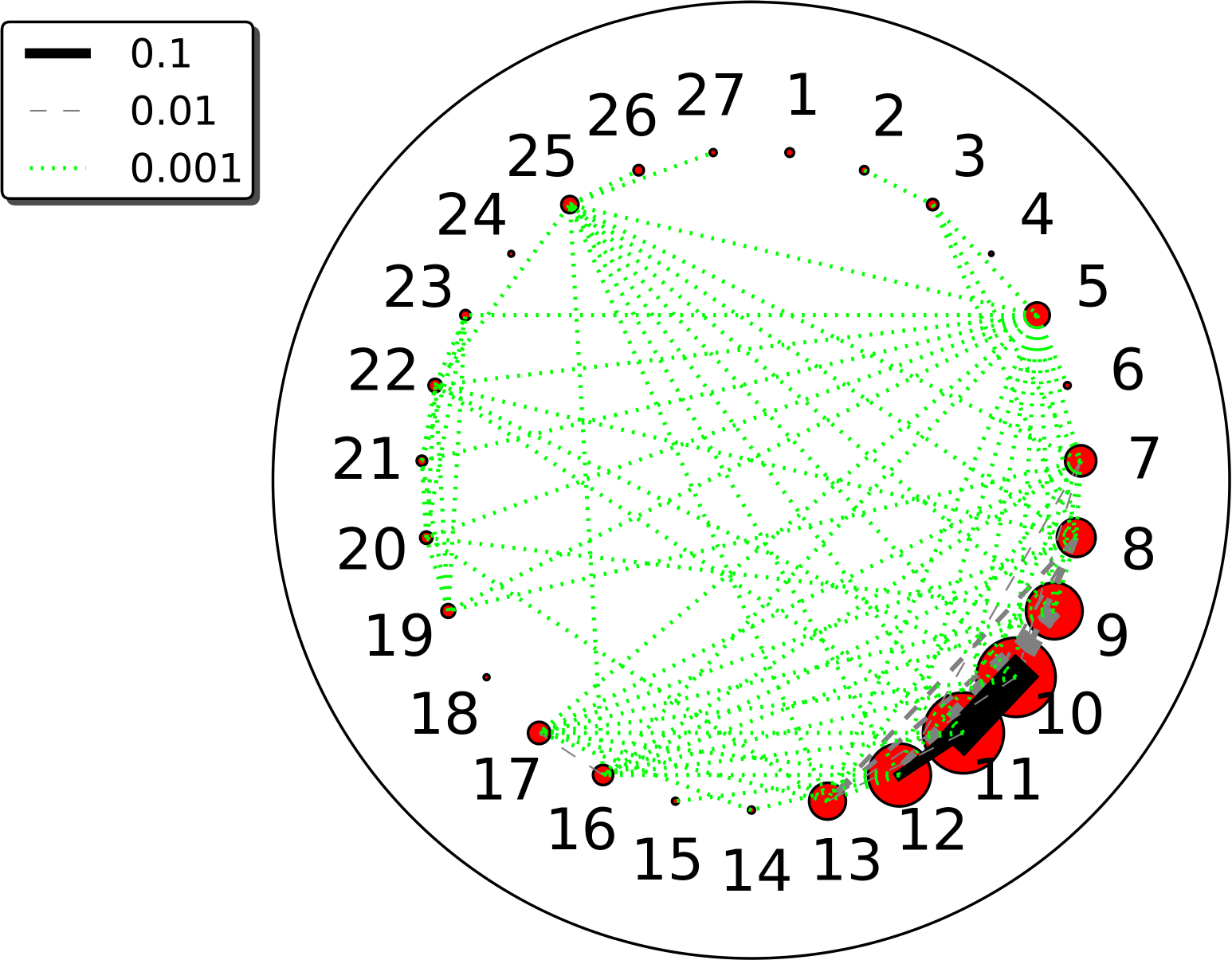} 
    \caption{Entanglement plots from PE-DMRG(20,27) calculations on the retinylidene chromophore (middle) showing the change upon 
vertical excitation from the $S_0$ ground state 
(left) to the first $S_1$ excited state (right). The magnitude of the single-orbital entropies are encoded in the size of the red circles while the 
magnitude of mutual information is ecoded in color and strength of the connecting lines. 
the thicker and darker the connecting line between two orbitals, the larger is their mutual information. 
The $\pi$ orbitals are labeled 5,7--10 and the $\pi^{*}$ orbitals 11--13,16,17.\label{mut_inf_gs_dmrg-20-27_m2p2}} 
\end{figure}

Before considering the shift of the $S_{1}\rightarrow S_0$ excitation energy due to the channelrhodopsin protein,  we analyzed the electron correlation 
mapped by the entanglement measures, i.e., by single-orbital entropies\cite{legeza2003a} and mutual information\cite{legeza2006,rissler2006}. 
These orbital-based entities have become popular as descriptors to classify 
multiconfigurational character.\cite{boguslawski2012b,boguslawski2014,stein16a,stein16b}
Figure \ref{mut_inf_gs_dmrg-20-27_m2p2} shows the PE-DMRG entanglement plots of the ground-state $S_0$ (left) and the first excited state $S_1$ (right). 
The entanglement plots indicate multiconfigurational character already in the ground-state with large mutual information particular between the $\pi$ and $\pi^{*}$ type orbitals 
(orbital numbers 5,7--13,16,17).  
Both single-orbital entropies and mutual information increase (as expected) for the first excited state. This emphasizes that for state-specific optimization 
of retinalidyne chromophores, multiconfigurational methods are required. 

The shifts of the $S_{1}\rightarrow S_0$ excitation energy caused by the protein in the PE-DMRG and PE-DMRG--srDFT calculations are 
compiled in Table \ref{pe-dmrg-srdft}. For comparison, we include in this table also results from the literature obtained with PE-TDDFT\cite{sneskov2013}, 
PE-CC2\cite{sneskov2013} and PE-TDCAS(6,6)--srDFT\cite{hedegaard2015a}. Before discussing the results, we should stress that PE-CC2, 
PE-TDDFT, and PE-TDCAS--srDFT are all based on response theory, and hence cover the effect of the environment in a fundamentally different way compared 
to state-specific PE-DMRG and PE-DMRG--srPBE. Furthermore, 
the PE-TDCAS(6,6)--srPBE method also includes orbital optimization whereas 
the present PE-DMRG and PE-DMRG--srPBE results are based on PE-HF or PE-HF--srPBE orbitals (for the ground state).     

\begin{table}
\caption{Shifts of the first $S_{0}\rightarrow S_{1}$ excitation of the retinalidyne chromophore due to the channelrhodopsin protein (in eV). The shifts are taken with 
respect to an isolated-molecule calculation of retinalidyne in its structure adopted within the protein. The strain exerted on the structure by the protein was previously found to have 
an effect of 0.05 eV on the $S_{0}\rightarrow S_{1}$ excitation\cite{sneskov2013}.     
We recall that 'TD' refers to linear-response calculations.
 \label{pe-dmrg-srdft}} 
\centering
 \begin{tabular}{lc}
\hline \hline \\[-1.5ex]
%--------------------------------------------------------------------------------------
Method                                         &  Shift     \\
\hline \\[-1.0ex ]
PE-DMRG[1024](20,27)                           &  1.01      \\[0.5ex] 
PE-DMRG[1024](20,27)--srPBE$^a$                & 0.52        \\[0.5ex]
PE-CC2\cite{sneskov2013}                       & 0.89    \\[0.5ex]
PE-TDDFT\cite{sneskov2013}                     & 0.49     \\[0.5ex]
PE-TDCAS(6,6)-srPBE\cite{hedegaard2015a}       & 0.69 \\[0.5ex]
\hline \\[0.5ex]
exp.\ \cite{ande05,niel06,ritt08,kato12}    & 0.58--0.70 \\[0.5ex]
%-------------------------------------------------------------------------------------
\hline \hline
 \end{tabular}

$^a$ short-range PBE functional as defined in Ref.~\citenum{goll2005}  \\ 
$^b$ short-range PBE functional as defined in Ref.~\citenum{fromager2007}
\end{table}

The excitation energy shift due to the protein calculated with PE-DMRG[1024](20,27) and PE-CC2 are similar, namely 1.01 eV and 0.89 eV, respectively.
However, PE-DMRG[1024](20,27) lacks most of the dynamic correlation and srPBE then reduces this shift to 0.52 eV, which is
comparable to the shifts calculated with TDDFT\cite{sneskov2013} 
(with the CAM-B3LYP\cite{yanai2004} functional) and TDCAS(6,6)--srPBE\cite{hedegaard2015a}.
We may compare the calculated shifts with experimental values of 0.58--0.70 eV taken from gas-phase \cite{ande05,niel06} and in-protein
\cite{ritt08,kato12} measurements, which show that almost all theoretical results listed in Table \ref{pe-dmrg-srdft} are in this range.
Only the pure PE-DMRG is slightly off, which indicates the importance of dynamic correlation. However, we also emphasize that
the orbitals were not optimized for the PE-DMRG wave function, which will be worst for the excited state. Moreover, we note that all
calculated results were obtained for a single structure and that was the one in the protein.

The absolute excitation energy from DMRG[1024](20,27) is overestimated in vacuum (3.17 eV), and also in the protein (4.18 eV). 
The experimental gas-phase peak maximum is between 2.00 eV and 2.03 eV \cite{ande05,niel06}, whereas it is shifted to 2.58 and 2.70 eV
in the protein \cite{ritt08,kato12}.
Only srDFT captures the neglected dynamical correlation in such a way that for DMRG--srPBE and PE-DMRG--srPBE we finally obtained 
improved excitation energies of 2.75 eV in vacuum and 3.27 eV in the protein, respectively. 
For comparison, CC2 and TDDFT yield excitation energies of 2.11 eV and 2.42 eV in vacuum and 3.0 eV and 2.91 eV in the protein, respectively.    
The PE-DMRG--srPBE result could be significantly improved by orbital optimization, because, as noted before, the HF-type orbitals in all DMRG 
calculations are certainly not adequate for the description of the excited state and hence lead to a larger deviation from experiment.

\section{Conclusion}\label{conclusion}

We presented a coupling of our MPO-based DMRG program with the Polarizable Embedding scheme that explicitly accounts for large 
environments using atom-centered multipoles and polarizabilities.
With this PE-DMRG scheme we investigated a well-studied system, namely the excitation energy and associated solvent shift of the 
 first excitation ($1^{1}A_1 \rightarrow 1^{1}B_1 $) of a water molecule embedded in a water environment. Ground and excited state energies 
were obtained from state-specific DMRG(10,30)[512] and PE-DMRG(10,30)[512] with an active space including all orbitals. 
We also compared PE-TDDFT and state-specific PE-DMRG with respect to the excitation energy and solvent shift for a range 
of embedding potentials of increasing accuracy. The applied potentials were denoted M0 (charges), 
M2 (charges, dipoles and quadrupoles), and M2P2 (charges, dipoles, quadrupoles, and polarizabilities). For a small environment, the 
addition of polarization is less important, whereas both the absolute excitation energy and solvent shift are affected significantly for a larger solvation shell. 
 The differences are somewhat larger for PE-TDDFT, compared to the state-specific PE-DMRG results. Further, the PE-TDDFT results show that the differences between
the M0, M2 and M2P2 PE potentials also depend on the density functional.   
The functional dependence is even more pronounced when the water molecule described by TDDFT has a stretched {O--H} bond. 

We also showed how the PE-DMRG scheme can be extended to a range-separated PE-DMRG--srDFT hybrid scheme\cite{hedegaard2015b} that accounts for dynamical correlation 
 when the QM system is too large to include all orbitals in the active space.  
With this method we investigated the blue-shift induced by a channelrhodopsin protein
on the $S_{0}\rightarrow S_{1}$ excitation of the protein's chromophore, 
a retinalidyne Schiff base. In accordance with previous studies, the protein effect was found to be large. The inclusion of dynamical 
correlation through a srDFT functional reduced the absolute excitation energy of $S_{0}\rightarrow S_{1}$ as compared to PE-DMRG 
(where they were considerably overestimated).   

It is currently not known to what extend incorporation of quantum effects in the interaction between QM region and PE potential will affect the results obtained here. 
This was done (for TDDFT) in a recent extension\cite{olsen2015a} of the PE scheme and these developments can be straightforwardly included in the method(s) presented here.  

%% classes.  Rather than use \section, an appropriate macro is
%% provided that will always work.
%%%%%%%%%%%%%%%%%%%%%%%%%%%%%%%%%%%%%%%%%%%%%%%%%%%%%%%%%%%%%%%%%%%%%
\section*{Acknowledgements}

E.D.H. thanks the Villum Kann Rasmussen Foundation for a postdoctoral fellowship. 
This work has been financially supported by ETH Z{\"u}rich and the Schweizer Nationalfonds (SNF project 200020\_156598).

%%\bibliographystyle{InorgChem}
%\bibliography{dmrg}
%%\newcommand{\Aa}[0]{Aa}

\newcommand{\Aa}[0]{Aa}
\providecommand{\refin}[1]{\\ \textbf{Referenced in:} #1}

\end{document}